# A novel sustainable role of compost as a universal protective substitute for fish, chicken, pig, and cattle, and its estimation by structural equation modeling

Hirokuni Miyamoto[1,2,3,4*#], Wataru Suda[2*#], Hiroaki Kodama[1], Hideyuki Takahashi[5], Yumiko Nakanishi[2], Shigeharu Moriya[6], Kana Adachi[7], Nao Kiriyama[7], Masaya Wada[8], Daisuke Sudo[9], Shunsuke Ito[10], Minami Shibata[10], Shinji Wada[10], Takako Murano[11], Hitoshi Taguchi[12], Chie Shindo[2], Arisa Tsuboi[1,3,4,6], Naoko Tsuji[4], Makiko Matsuura[1,4], Chitose Ishii[2,4], Teruno Nakaguma[1,3], Toshiyuki Ito[14], Toru Okada[15], Teruo Matsushita[3,4], Takashi Satoh[16], Tamotsu Kato[2], Atsushi Kurotani[6], Hideaki Shima[6], Yudai Inabu[5], Haruki Yamano[5], Yukihiro Tashiro[17], Kenji Sakai[17], Kenichi Mori[1,3,4], Kenta Suzuki[21], Takeshi Miura[13], Hidetoshi Morita[18], Shinji Fukuda[1,19,20], Jun Kikuchi[6], Hisashi Miyamoto[4,22], Masahira Hattori[2,23], Naoki Yamamoto[24], and Hiroshi Ohno[2,20*]

*Affiliations:*

*1. Graduate School of Horticulture, Chiba University, Matsudo, Chiba 271-8501, Japan*

*2. RIKEN Center for Integrative Medical Sciences, Yokohama, Kanagawa 230-0045, Japan*

*3. Japan Eco-science (Nikkan Kagaku) Co. Ltd., Chiba, Chiba 260-0034, Japan*

*4. Sermas Co., Ltd., Ichikawa, Chiba 272-0033, Japan*

*5. Kuju Agricultural Research Center, Kyushu University, Takeda, Oita 878-0201, Japan*

*6. RIKEN Center for Sustainable Resource Science, Yokohamai, Kanagawa 230-0045, Japan*

*7. Crest Co., ltd., Komakishi, Aichi 485-0802, Japan*

*8. Wada farm Co.ltd., Gujou-shi, Gifu 501-5302, Japan*

*9. Nihon Layer, Sano, Gifu 501-1101, Japan*

*10. Chuubushiryo Co. Ltd., Oobu, Aichi 272-0033, Japan*

*11. Chiba Prefectural Livestock Research Center, Yachimata, Chiba 289-1113, Japan*

*12. Hirano Co.,Ltd., Narita, Chiba 287-0217, Japan*

*13. Southern Ehime Fisheries Research Center, Ehime University, Ainan, Ehime 798-4292, Japan*

*14. Keiyo gas energy solution Co. Ltd., Ichikawa, Chiba 272-0033, Japan*

*15. Asuka Animal Health Co., Ltd.,Tokyo 108-0023, Japan*

*16. Kitasato University, Graduate School of Medical Sciences, Sagamihara, Kanagawa 252-0373, Japan*

*17. Center for International Education and Research of Agriculture, Faculty of Agriculture, Kyushu University, Nishi-ku, Fukuoka 819-0395, Japan*

*18. Okayama University, Okayama, Okayama700-8530, Japan*

*19. Institute for Advanced Biosciences, Keio University, Tsuruoka, Yamagata 997-0035, Japan.*

*20. Intestinal Microbiota Project, Kanagawa Institute of Industrial Science and Technology, Ebina, Kanagawa 243-0435, Japan.*

21. RIKEN, BioResource Research Center, Tsukuba, Ibaraki 305-0074, Japan

*22. Miroku Co. Ltd., Kitsuki, Oita 873-0021, Japan*

*23. School of Advanced Science and Engineering, Waseda University, Tokyo169-8555, Japan*

*24. National Center for Global Health and Medicine, Ichikawa 272-8516, Japan*

# Co-first Author

* Co-correspondence: Hirokuni Miyamoto Ph.D., Chiba University, RIKEN, and Sermas Co. Ltd.

E-mail: hirokuni.miyamoto@riken.jp, h-miyamoto@faculty.chiba-u.jp

Co-correspondence: Wataru Suda Ph.D., RIKEN

E-mail: wataru.suda@riken.jp

Co-correspondence: Hiroshi Ohno M.D. & Ph.D., RIKEN

E-mail: hiroshi.ohno@riken.jp

## Abstract

**Natural decomposition of organic matter is essential in food systems, and compost is used worldwide as an organic fermented fertilizer. However, as a feature of the ecosystem, its effects on the animals are poorly understood. Here we show that oral administration of compost and/or its derived thermophilic Bacillaceae, *i.e., Caldibacillus hisashii* and *Weizmannia coagulans*, can modulate the prophylactic activities of various industrial animals. The fecal omics analyses in the modulatory process showed an improving trend dependent upon animal species, environmental conditions, and administration. However, structural equation modeling (SEM) estimated the grouping candidates of bacteria and metabolites as standard key components beyond the animal species. In particular, SEM model implied a strong relationship among partly digesting fecal amino acids, increasing genus *Lactobacillus* as inhabitant beneficial bacteria and 2-aminoisobutyric acid (AIB) involved in lantibiotics. These results highlight the potential role of compost for sustainable protective control in agriculture, fishery, and livestock industries.**

***Running title***: Compost candidate protective for animals

## Introduction

Tilman D. et al. suggested in 2001 that new incentives and policies for ensuring the sustainability of agriculture and ecosystem services are crucial[1]. The planet boundaries framework[2] defined safety for humanity and the stability of the Earth system. In particular, improving the nitrogen cycle[3,4] and biodiversity[5-8] are urgent issues in the agriculture, fishery, and livestock industries. The driving force of climate change and the consumer importance of water demand is also concern[9]. Since excessive utilization of artificial agricultural chemicals and antibiotics may promote the occurrence of agricultural drug- or antibiotic-resistant pathogens[5], the global action plan against antimicrobial resistance (AMR) is an essential element of the sustainable development goals (SDGs). Therefore, concrete techniques[10,11] and policies[12,13] for sustainable development are increasingly necessary, as is communicating the seriousness of these issues to a wide range of people [1].

This study should provide a novel perspective for considering these issues from resource recycling. The need for a sustainable food system is a global issue. Generally, compost is fermented organic matter, and several types of organic raw materials, *i.e.,* plant residues, food waste, animal feces, and sludge, can be used to produce compost. Compost is used as an organic fertilizer in organic agriculture. All animals have indirectly ingested the microorganisms in compost via botanical food, feeds, and fish after soil flows into the river and sea. Nevertheless, the roles of compost in ecosystems are poorly understood. The qualities of composts seem to be unstable since several types of raw materials are fermented under ambiguous fermentation conditions characterized by bacterial fermentation species, moisture, and other conditions of the fermentation process. However, we have reported a compost model of the fermented marine animal resources in a fed-batch system of bioreactors by a stable thermophilic bacterial community[14]. The compost was predominantly composed of thermophilic Bacillaceae bacteria with antifungal activity[14]. The compost amendment to soils promotes plant growth together with denitrification activities[15]. The amendment to soils and wood chips for habitat of beetle larvae tend to reduce the phylum Gemmatimonadetes[16,17], a bacterium with anammox (anaerobic ammonium oxidation) reaction, to suppress the generation of $N_2O$, which has a high global warming coefficient (approximately 300 times worse than $CO_2$). The oral administration of the compost or its extract improved the fecundity and quality of flatfish[18], carp[19], chickens[20], and pigs[20,21]. As a result of many years of research, it has been confirmed that the seagrass *Zostera marina*, which is one of the protagonists for blue carbon (absorption up to 40 times of $CO_2$ more efficient than land forest soils), grows only around the fishery farming facility that was fed the compost[22]. The gut immune response and antioxidant activity in rodent models were raised [23,24]. An experiment with gnotobiots administered the compost extract was performed to explore the functional thermophiles in the compost. Two types of thermophiles, strains closely related to the 16SrRNA sequences of *Bacillus thermoamylovous* and *Bacillus coagulans*, were isolated from the compost. These strains raised immune activities in the conventional mice [33]. Then, the strain closely related to *Bacillus thermoamylovorans* [33] was registered as *B. hisashii* [25], and recently reclassified to *Caldibacillus hisashii*[26]. The strain closely related to *Bacillus coagulans* also reclassified to *Weizmannia coagulans*[26]. The effects of these single bacteria on livestock animals have also been evaluated, and the administration has obtained remarkable discoveries for cattle. The oral administration of *C.hisashii* to calves enhances their growth together with a rise of the phylum Bacteroidetes[27]. It reduces the fecal population of the genus *Metanobrevibacter* sp. involved in the generation of methane gas, which has a high global warming coefficient (approximately 25 times worse than $CO_2$). These observations suggest that this compost and its derived thermophiles consistently exert various functions valuable for animal and plant health and environmental conservation. However, its mechanism of action is not well understood. Here, the aim of our study is to assess host-microbiota interactions in animals after exposure to the compost and/or the thermophile itself. Therefore, the productivities and qualities of several animals after applying the compost were analyzed. In fish, chickens, pigs, and cattle, the effects of oral administration of the compost exposure and/or their derived Bacillaceae and their productivity, physiological outputs, and fecal bacterial flora were surveyed. An interesting trend was observed with utilizing a thermophile as a feed additive: changes in the diversity of intestinal flora; reduced relative death rates; reduced antibiotic use, etc. Furthermore, the relationships between the protective effects of this compost, fecal microbiota, and metabolites were estimated by structural equation modeling according to association analysis. This study is the first report to evaluate the importance of compost quality in the environmental cycle across animal species to the best of our knowledge. Furthermore, the compost and the derived thermophiles used in this study is expected as the substitute of antibiotics, which will contribute to biodiversity, and as a feed additive efficiently digest excess feed, which reduce the nitrogen load in the external environment. Thus, this study provide a novel perspective for sustainable food production, recycling, and ecosystem.

## Results

### *The difference in the fecal microbiota between the animals*

The difference in the fecal bacterial communities between the animals is shown in Fig. S1. In our selected samples, the values of diversity were dependent upon the animal species (Figs. S1a, S1b, and S1c), and the phylum Firmicutes was abundant in all the animals (Fig. S1d). The phyla Rhodophyta and Proteobacteria in fish and Bacteroidetes in chickens were the second most abundant. Several genera were detected in fish, but genera *Bacillus* and *Ralstonia* were detected at slightly higher levels (approximately 10% relative abundances in fish feces) (Fig. S1e). In the land animals, lactic acid bacterial genera *Lactobacillus* and/or *Streptococcus* were predominant (Fig. S1e). Such trends appeared to be dependent upon animal specificities. Thus, the physiological effects of compost were evaluated in each animal.

***Evaluation in the fish model***

In flatfish farms with the application of compost or its extract, the death rate of flatfish is reportedly reduced[6]. One of the causes of flatfish death is Edwardsiellosis, which is known as an infectious disease caused by *Edwadsiella tarda*, a widely transmitted fish pathogen. Therefore, the defensive effects of the compost on red sea bream bred in the laboratory tanks were examined after the artificial infection with *E. tarda* in the two regions, Shizuoka and Ehime, in Japan (Fig. S2a). After long-term pre-administration of the compost extract (FS-Test), the phenotype such as the weight of the fish was normal (Extended fig. 1a). After intraperitoneal injection with *E. tarda*, the body surfaces differed among the groups of fish (Extended fig. 1b), and the survival of the fish in the compost group was little decreased (Extended fig. 1c). Interestingly, recovery after anesthesia appeared to be improved, although not significanly (Fig. S2b). Previous studies have reported that rodents orally administered with the compost extract exhibit increased fecal IgA[24,28], but IgA is undetectable in fish[29]. Therefore, serum complement activity in the fish was examined as an indicator of immune system activation. The serum complement activity just after short-term pre-administration of compost during 70 days (Day 0, just before the injection) was increased in the compost group ($p = 0.007$ and $p= 0.011$)(Extended fig. 1d). Additionally, after short-term pre-administration of compost (FE-Test) (Fig. S2c), the survival in the compost group improved (Extended fig. 1e). To explore this finding, organic acids in the intestine were examined. No significant changes were observed, although the succinic acid concentration in 5%-compost group was low but tended to decrease ($p=0.04$) (Fig. S2d). The diversities of the gut bacterial communities were analyzed by sequencing 16SrRNA genes from the fecal samples; the results indicated moderate increased difference in diversities, not significantly (Extended fig.1f)(Fig. S2ef). The differences in phylum populations in the comparison groups were not significant (Fig. S2g). Still, the following bacteria significantly differed among the groups (Extended fig. 1g and Fig. S2h): OTU00017 *Clostridium perfringens*, which is a pathogenic bacterium[30], OTU00014 *Exiguobacterium* sp., which can adapt to changes in pH, salinity, and UV radiation[31], OTU00018 *Salinicoccus jeotgali*, which is a halophile[32], OTU00027 protobacterium symbiont of *Nilaparvata lugens*, which is a pest of rice[33], were significantly decreased in the FE-5% compost group, and *Lactococcus* sp. were significantly increased in the compost groups. Interestingly one species of *Lactococcus* sp. has been reported as a probiotic bacterial candidate with antipathogenic activities[34,35]. Thus, the increment of fecal bacterial diversities and/or decreased population of infectious bacteria in the compost treatment appeared to be associated with the survival of the infectious ones. The population analysis of thermostable bacteria in the feces showed that *Caldibacillus thermoamylovoras* and/or *C. hisashii* (synonym: *Bacillus thermoamylovoras* and *C.hisashii*)[26] and *Weizmannia coagulans* (synonym: *B. coagulans* )[26] were abundant in the compost groups (Extended fig.1h). Thus such dynamics in the gut environment of fish exposed to the compost could aid in fish survival against *E. tarda* infection. In particular, *C.hisashii* and *W.coagulans* may be key bacteria in the compost.

***Evaluation in the chicken model***

The growth effects of the compost extract on chicks were examined (CC-Test in Fig. S3a). The growth rates significantly increased after 9 weeks by the continuous compost exposure (Extended fig. 2a). Randomly selected seven healthy chicks were fed the compost for 13 weeks, the untreated and treated chicks were injected with *Salmonella* sp., respectively (CC Test in Fig. S3a). In the ovary tract of chicks with compost extract, *Salmonella* sp. was rarely detected, although it was seen in the cecum (Extended fig. 2b). The diversities of the fecal microbiota slightly appeared to alter one another (Fig.S3bc). Especially the phylum Actinobacteria had an increased tendency ($p=0.16$), and phylum Firmicutes, the genera *Enterococcus* and *Kurthia* markedly decreased ($p=0.041$, $p=0.091$, and $p=0.008$, respectively) (Extended fig. 2c and Fig. S3de). On the other farm, which was always maintained as a production farm, the effects of the compost extract with a minimum concentration of *C.hisashii* N11 strain and *W.coagulans* N16 strain as probiotic candidates [24], which were also identified by sequencing in feces of red seabream in this study, were investigated (CW Test in Fig.S3a). Two groups were prepared in a newly built henhouse: One hen group was fed the treatment with *C.hisashii* and *W.coagulans*, while a second hen group was not treated, respectively. At 3 months with feeding, the fecal bacterial diversities of the healthy hens showed a significant difference within the same newly house (Fig.S3fg). In particular, genus *Kurthia* had a decreased tendency ($p=0.133$), and genus *Lactobacillus* was significantly increased ($p=0.018$) (Extended fig. 2d): and these results were similar to those of the CC test. Although all the short-chain fatty acids (SCFAs), which modulate gut physiological function[36-38], did not always increase, the fecal propionate level had an increased tendency at 1 month after the compost exposure ($p < 0.2$) (Extended fig. 2e), consistent with the slight rise of fecal IgA content ($p=0.25$)(Extended fig. 2f). Since the conditions of CW Test did not cause infectious disease, new tests were conducted at the hen houses where infectious disease, necrotizing enterocolitis with *Clostridium* and the related bacteria, sometimes occurred (CCR test in Fig.S3a). In these houses, necrotizing enterocolitis occurs in the summer. When the temperature in the hen houses was extremely high, the death rate in the compost administered hen house was similar to that in the negative control hen house maintained under moderate temperature (Extended fig.2fg). The correlation of temperature in the summer and the death rate per week yielded the difference by the compost exposure. However, the fecal bacterial diversities were significantly different (Fig.S4de), the increment of genus *Lactobacillus* was monitored as described in the chicken tests above (Extended fig.2h). The population of the genus *Veillonella*, which enhances athletic performance by studies of the human gut microbiome[39], increased in the minor bacterial communities (Fig.S4f). The possibility that chicken with necrotizing enterocolitis caused by the genus *Clostridium* and that the compost extract administration reduced the related microbes was speculated. These effects can be explained by changes in intestinal flora caused by the administration of compost and concomitant changes in intestinal metabolites such as SCFAs, carbohydrates, and amino acids.

***Evaluation in the pig model***

Two tests were utilized to explore the key probiotic bacteria in the compost. In the tests of two swine farms (PK and PF Test in Fig.S5a), *C. hisashii* as a probiotic candidate, which was more efficient than the other isolates in the previous gnotobiote test[24], was used. The bacterial diversities in the pigs' feces differed between the two environments (Fig.S5bc). The diversities in the administered pigs were weakly different (Fig.S5d). In both short-term tests, a decrease in the population of the genus *Streptococcus* was commonly observed after the administration (Extended fig.3a). Based on these observations in the farm controlled for the experiment, the experiments in the other two swine production farms were prepared, and the effects of oral administration of *C.hisashii* with the compost extract (PI-Test) and without the compost (PM-Test) on the fecundity of pigs were estimated (Fig.S5a). In PI-Test, the death rate of growing pigs decreased to approximately one-fifth after continuous administration, despite the highly reduced usage rates of four antibiotics per individual pig (Extended fig.3b). Under these conditions, the levels of IgA and some SCFAs at the final stage of the growing stage increased (Fig.S6ab). The difference in fecal bacterial diversities and phyla was insignificant (Fig.S6cde). However, the genus *Lactobacillus* significantly increased after administration, and *Streptococcus* decreased (Fig.S6f).

In the PM-Test, the fecal IgA in the piglets was significantly increased by the *C.hisashii* administration, but only one exception was observed (piglets from sow No.1808) (Fig.S6g). After administration, the death rates of growing pigs were significantly at low levels (3.56%, 3.455, and 1.86%) (Extended fig.3d). The cost of the drugs used during the same period also substantially decreased (307, 283, and 254 yen per individual pig) (Extended fig.3e). To clarify the effects of the *C.hisashii* on the fecal bacterial population, the difference between males and females was examined on another occasion. An increment of fecal IgA (Extended fig.3f and Fig.S7a) was associated with the production of SCFAs (Extended fig.3g and Fig.S7b). However, the increase in IgA in the females appeared to be remarkable (Fig.S7a). Under these conditions, an increase in the phyla Bacteroidetes and Actinobacteria and a decrease in Firmicutes were observed (Extended fig.3h), although the fecal bacterial diversities in male and female pigs differed slightly. Thus, the gut flora of pigs could be improved by the compost-derived thermophile, reducing the use of synthetic drugs.

***Evaluation in the cattle model***

Since animal proteins derived from marine animals as raw materials before fermentation may be included in compost, they cannot be legally used as feed additives for cattle. Therefore, only *C.hisashii* was implemented in these tests. In addition, since a mother cow lays only one calf, which is expensive, it is impossible to set up various test systems, such as in pigs. Therefore, we analyzed some of the fecal conditions of the little six calves (Fig. S4ab). In young calves with diarrhea (Extended fig.4a), the recovery tendency of diarrhea was confirmed after the administration of *C.hisashii* (Extended fig.4b). The $H_2O$ ratio appeared to be decreased (Fig. S8c). The fecal IgA contents increased after the administration (Extended fig.4c), concomitant with increased SCFA levels (Extended fig.4de). The fecal $H_2O$ content was inversely correlated with the SCFA content (Extended fig.4f). Under such conditions, fecal bacterial diversities did not differ before and after the administration. However, genus Fusobacterium had a decreasing tendency after administration (Extended figs.4g and 4h). The detection of pathogenic genes of *Escherichia coli* clarified that some of the tested calves were positive for sequences of Shiga toxin genes (No.1294, No.1293, 1295) (Fig. S8g). However, the causes of diarrhea in the other individuals were not clarified. Thus, these observations suggest that calves' pathogenic and/or non-pathogenic diarrhea could be improved through alteration of the fecal bacterial population by the specific thermophile administration.

***Prediction of features beyond the animal species***

Here, it was confirmed that the administration of compost and/or thermophile activated physiological protection in sea and land animals, even though the changes in intestinal flora were specific to each animal species. These observations have assumed an association between changes in fecal bacterial communities, their metabolites, and protective action. However, the manner of this change is not constant and varies depending on the animal species. Therefore, comprehensive metabolome analyses were performed on typical animal species in which samples exist (Fig.S9-S10). As a result, the correlation of metabolites beyond the animal species was confirmed (Fig. S11), and the relative differences dependent upon the animal species were occurred by administration of compost and/or a thermophile (Fig.S12). Pathway enrichment analysis identified the metabolic pathways in common among the animals and those that were different. An association analysis, as an unsupervised machine learning method, was conducted to explore universal factors that administered compost across animal species (Fig. S13). Interactive components standard in the animals treated with compost and/or thermophile were clarified by association analysis. Of the associated metabolites, AIB, butyrate, and *Lactbacillus* was selected as increased components (Fig. S13). The free amino acids and amino acid-related metabolites in the feces were selected as decreased components. These increases and decreases for each animal species are shown in Fig. S14. Further energy landscape analysis, as an unsupervised machine learning method, examined the effects of composting across species. As shown in Fig. 2a, the macroscopic effects of composting were not discernible. On the other hand, it was suggested that there are differences at the factor and module levels (Fig. S15). Therefore, association analysis and energy landscape analysis were used to narrow down the factors assumed to be effective for composting, and structural equations including AIB, butyrate, and *Lactobacillus* were calculated, focusing especially on the factors that increase by administration of compost or *Caldibacillus*. After evaluating the regression equations in Tables 1 and 2, the optimal structure models (SEM) were estimated (Figs. 2b and 2c). Of these components, glutamic acid and tryptamine were the factors that tended to decrease with composting (Fig. 2d). In addition, candidate factors causally related to these increasing AIB, butyrate, and *Lactobacillus* were evaluated by Markov networks (Fig. S16). It is well known that *Lactobacillus* and butyrate are effective in the intestine. However, little effective information on the intestine is known about AIB. Therefore, based on the previous reports[40,41] on AIB, we were able to assume the effect of AIB on pathogenic microorganisms. First, the compost and the supernatant of the *Caldibacillus* culture contained trace amounts of AIB. Next, the direct effect of compost itself on pathogenic microorganisms was investigated. The results showed no antibiotic effect (data not shown). Therefore, its effects on viruses were investigated (Fig. 3). The results showed that it had an inhibitory effect on infectivity against RNA viruses: *Alphacoronavirus*, *Betacoronavirus*, and *Influenzavirus* (Fig. 3a). These effects were also confirmed in the experiment with *Caldibacillus*, a thermophile that is the core of the compost (Fig. 3b). Interestingly, the effect was suggested to have an increasing trend in culture supernatants. *Caldibacillus* was administered to germ free mice and the concentration of AIB in the cecum was detected. The results showed that the AIB concentration tended to increase when vegetative cell and culture supernatant to the mice were included (Fig. 3c), although at low levels in the mice administered the spore of *Caldibacillus* (Fig. 3d) and similar trends on changes in other fecal metabolites (Fig. S17).

## Discussion

Here, it was speculated that physiological protection of fish, chickens, pigs, and cattle commonly tends to improve by oral administration of compost and/or thermophile, despite alteration of their gut flora depending on the animal species and their habitat. As a result of pathway analysis to find a common mechanism beyond the animal species, some metabolic pathways common to animal species were observed. Further statistical structure equation modeling (SEM) based on selected data using machine learning as association analysis and energy landscape analysis estimated *Lactobacillus,* butyrate, and 2-aminoisobutyrate (AIB) as component candidates optimally linked with the compost beyond species. *Lactobacillus* is widely known to play a role in the intestine as one of the lactic acid bacteria[42,43]. Butyrate are produced by commensal bacteria and promote peripheral regulatory T cell generation[44]. AIB is a molecule known to be involved in the synthesis of alamethicin and/or lantibiotic, which may have antibacterial activities[45,46]. These factors may play a role in the effect of the compost, although the molecular mechanism of the compost and/or thermophile that improves the intestinal flora is unknown. As one related piece of information, a rise in butyrate under the moderate temperature in artificial anaerobic conditions associated with the compost administration[47] is noted. The improvements of the animal's healthy balances by the compost and/or thermophile may be related to the physiological roles of SCFAs, which have been shown to protect the gut function in mouse models[36,38].

Furthermore, it has recently been suggested that AIBs can synthesize lantibiotic, which play a role in increasing cell permeability[40]. Papain is involved in these synthesis[41]. Interestingly, *Caldibacillus* had papain-like protease and lantibiotic Sun A-related gene in the genome. Also, some of the lantibiotic and cyclic peptide are known to be effectively disrupt against viral infectivity[48-50]. Therefore, the suppression effect against viral infectivity was evaluated and the effects was confirmed (Fig. 3).

These data may point to the need for further research on RNA virus control in the gut. The involvement of bacteriophages and intestinal flora is obvious[51,52]. However, the study of RNA bacteriophages in the gut is still unexplored. In this study, a trend toward less incidence of diarrhea due to PED infection was observed in compost-added swine housing (data not shown). In addition, a trend toward an increase in *Lactococcus* in fish trials has been observed in compost-treated groups (Extended fig. 1g). Interestingly, some RNA bacteriophages are known to target *Lactococcus*[53,54]. Although these relationship between intestinal bacteria and the RNA bacteriophage were not examined in this study, it would be an interesting hypothesis if RNA bacteriophages in the gut of fish were responsible for the increase in *Lactococcus* caused after oral administration of the compost. Based on these results, in addition to Fig. 1c, the effect of compost is hypothesized in Fig. 4. Compost-derived *Caldibacillus hisasshii* is involved in the synthesis of AIB and lantibiotic including AIB in the gut. It is hypothesized that these and other factors are intertwined to improve the intestinal microbiota. One factor supporting this hypothesis is that AIB actually tends to increase in germ free mice after oral administration of *Caldibacillus hisasshii* (Fig. 3c). Although papain like protease is present in the genome sequence, the results of structural analysis were not always suitable. The predicted peptide structure (Fig. S18) was not calculated as a structure with high accuracy. On the other hand, there were known cases where the RNA virus itself uses its own papain like protease for infection[55]. As one of the possibilities, these facts should be considered in relation to the gut microbiota, such as the effects of competitive inhibition due to the presence of enzymes similar to papain.

In addition, quorum sensing by the genus *Bacillus* has been shown to depress the population of staphylococci[56]. The possibility of a similar mechanism and an additional mechanism is speculated in this study. Previous data indicate that one mechanism may involve quorum sensing by a strain of *B. subtilis*, but it is often already included as a standard feed additive in the fishery and livestock industries of Japan. Therefore, the results of this study may indicate the independent effects of *C.hisashii*, although the sensing mechanism was not observed here.

The molecular mechanism should be investigated in future studies. Generally, there are several types of compost, and compost is considered an organic fertilizer that only affects crops. In all the environments, namely, those of soil and sea, the bacterial diversity is important for the health of humans and animals[6,7] because the bacteria in these environments are indirectly administered orally to various animals. Therefore, the kind of compost fermentation microorganisms should be an essential consideration. This study is the first report to evaluate the importance of compost quality in the environmental cycle across animal species such as sea and land animals to the best of our knowledge. Since these microorganisms can circulate in the environment, research of this sort is necessary for sustainability efforts and for improving nitrogen cycle[3,4] and biodiversity[5-8] with the ultimate goal of establishing high-quality, society-level recycling strategies.

## Materials and Methods

These information are described later as the *Supplementary Methods*.

**Acknowledgments** This research was partly supported by a Grant-in-Aid from the Chiba Prefectural Federation of Small Business Association, based on the foundation for manufacturing (Mono-Zukuri) of the Ministry of Economy, Trade, and Industry in Japan (Grant ID 25121200402). We are grateful to Masaya Wada (Wada Farm Co., Ltd.), Keisuke Kogusa, Yasuyuki Ishikawa (Chuubushiryo Co. Ltd.), Youta Aoi, Toshihito Shinmyo, Toshiyuki Ito, Motoaki Udagawa,Yu Kajihara, (Keiyo gas energy solution Co.,Ltd.), Jirou Matsumoto, Kazuo Ogawa (Sermas Co.,Ltd.), Kosuke Nakagiri (Kyushu University) for breeding facilities, sampling, and/or preparation of feed. Special thanks to Atsushi Ido (Ehime University), for special technical supports. This study was partly supported by a Grant-in-Aid from the Chiba City Foundation for the Promotion of Industry, a Grant-in-Aid from the Chiba Prefectural Federation of Small Business Association, based on the foundation for manufacturing (Mono-Zukuri) of the Ministry of Economy, Trade, and Industry in Japan (ID 25121200402), a Grant-Aid for Research and Development Support Program for regional revitalization of the food industry sector of the Ministry of Agriculture, Forestry, and Fisheries of Japan (No. 22331219048), the Ministry of Economy, Trade and Industry of Japan (No.21231219119).

**Animal ethics** All treatments of the animals were conducted in accordance with the institutional animal care guidelines for the farms. The management was according to individual farm-specific guidelines for most efficiently producing animals.

**Author contributions** Hirokuni Miyamoto, Kenichi Mori, Hisashi Miyamoto conceived and designed the experiments; Shunsuke Ito, Wada, Nakajima, and Takeshi Miura performed fish experiments.; Hirokuni Miyamoto, Youta Aoi, Nao Kiriyama, Kana Adachi, Daisuke Sudo, Takako Murano performed the experiments on chicken.; Hirokuni Miyamoto, Kenichi Mori, Minami Shibata, Shinji Wada, Toru Okada performed pig experiments. ; Hirokuni Miyamoto, and Hideyuki Takahashi performed the experiments of bovine.; Hirokuni Miyamoto, Makiko Matsuura, Chitose Ishii, Naoko Tsuji, Teruno Nakaguma, Hidetoshi Morita performed the biological assay.; Hirokuni Miyamoto, Wataru Suda, Yumiko Nakanishi, Tamotsu Kato, Arisa Tsuboi, Naoko Tsuji, Chie Shindo, Teruo Matsushita, Shigeharu Moriya, Atsushi Kurotani, Hideaki Shima, Jun Kikuchi, and Masahira Hattori, Hiroshi Ohno analyzed the raw data.; Hirokuni Miyamoto and Wataru Suda wrote the manuscript. All authors read and approved the final manuscript.

**Data availability** Raw files of the bacterial V1–V2 16S rRNA data are deposited in the DNA Data Bank of Japan (DDBJ) under NCBI Bio-Project accession numbers PRJDB9535 (PSUB012136). The 16S rDNA library in Extended fig.1 is also deposited in the DDBJ, under NCBI Bio-Project accession numbers LC646904-LC646942. In addition, the R protocols for association analysis used in this study were deposited on the following websites: http://dmar.riken.jp/Rscripts/ and http://dmar.riken.jp/NMRinformatics/.

**Competing interests** The authors declare no competing interests

**Figure legends**

Fig. 1 (a) the experimental design in this study. The compost fermented with unutilized marine animal resources and/or its derived probiotic candidate thermophile itself cultivated by bioreactor were administrated to fish, chicken (hen), pig, and cattle. Each animal species is evaluated under different conditions. For each group, fish were estimated at n = 33-88 (individual data), chickens at n = 27-58 (hen houses with more than 10,000 hens per henhouse), pigs at n = 4-62 (swine houses with more than 300 individuals per swine breeding facility), and cattle at n = 4 (individual data as an improvement evaluation for diarrhea cattle). Details in the method section is described. (b) shows the effects of a compost fermented with unutilized marine animals resources and/or their-derived thermophile on the relative accident rates of the various animals. (c) shows a hypothesis of the molecular mechanism of gut environments altered by compost and/or its derived thermophile, which were predicted by omics evaluation for each animal.

Fig.2 (a) The entire energy landscape linked with the (a) compost used in this study. The axis formed the energy landscape with compositional energy, community state, and compost. (b) The interaction network of Lactobacillus and AIB were shown, respectively. The color means as follows: Red, putative positive interaction; Blue, putative negative interaction. (b)(c) The relationship of fecal microbiota and metabolites was shown by structural equation modeling (SEM) calculated for groups selected by association analysis and ELA. Standardized β coefficients are reported. The abbreviation shows as follows: AIB, 2-aminoisobutyric acid; Butyrate, Butyric acid; Glu, glutamic acid; Lct, Lactobacillus ; Tst, Test. Green positive; purple, negative. The models and fit indices were shown in Table S2. (e) Relative abundance of genus *Lactobacillus* and relative ratio of 2-aminoisobutyric acid (AIB), Butyrate, and Glu to the control data was assessed for each animal.

Fig. 3 Inhibitory effect of thermophilic fermentation solutions against RNA viruses and estimation of their mechanism of action.

(a) Among RNAviruses with envelopes, PED (Porcine Epidemic Diarrhea) virus, an Alphacoronavirus, and SARS-CoV-2, a Betacoronavirus, and influenza virus, as well as The inhibitory effect of thermophilic fermentation solutions on infection with *Calicivirus*, a RNA virus without envelope, was evaluated for each cell. The virus strains and cell lines evaluated are indicated in the figure. blue line, untreatment group; dotted line, fermentation solution-treated group. (b) Inhibitory effect of *Caldibacillus hisashii* culture medium and viable cells against RNA viruses was tested; PED virus, an Alphacoronavirus, was used as the RNA virus. (c) Concentration of AIB in cecal feces of germ free mice treated with *Caldibacillus hisashii*, showing the results when vegetative cell *Caldibacillus hisashii* were administered during long term (151 days) (left) and when they were administered during short term (35 days) (right).

Fig.4 A hypothesis based on the present results is presented: it is likely that *Caldibacillus hisashii* itself acts positively in the biosynthesis of AIB. It also possesses a gene related to papain, which is required for AIB to be metabolized into the lantibiotic. In addition, it also possesses other lantibiotic genes in its genome. It is assumed that these may function and affect bacteriophages/viruses in the gut. It is possible that the gut microbiome could be affected after acting on bacteriophages/viruses.

**Extended figure legends**

Extended fig.1.

(a) The physiological conditions of red breams in the FS-test in Shizuoka (Fig.S.2a) were shown. (b) Photographs of sea breams administration with FS-control (condition without administration of compost extract), FS-1% compost extract, FS-5% compost extract at one month after the injection of *Edwadsiella tarda.* (c) shows survival rates of FS-test (n=10; p=0.07 in long rank test). (d) shows the complement activities before and after the injection (n=5). (e) shows the survival rates in *Edwadsiella tarda* exposed test as the repetitive test in Ehime (FE-test), In sea breams administration with blank (normal condition without the injection of *Edwadsiella tarda*), FE-control (condition without administration of compost extract), FE-1% compost extract, E-1% compost, and FE-5% compost, death rates of the fish were shown after the injection of *Edwadsiella tarda* (Blank, n=11; FE-control, n=30; FE-extract 1%, n=30; FE-1% compost, n=27; FE-5% compost, n=30; p=0.007 in long rank test). (f) shows α-diversity of the fecal bacterial community of fish tested in the FE-test. (g) shows the characteristic bacterial population (n=6; > 1% as average of relative abundance of OTU assigned sequences; p < 0.05 as FE-1% compost vs FE-control). The percentages show the identities. OTU00027 shows proteobacterium symbiont of *Nilaparvata lugens*. (h) shows characteristic thermostable bacterial species of the fecal microbiota of FE-control (fish without the compost exposure and with the injection of *Edwadsiella tarda*) and FE-5% compost (fish with the compost exposure and with the injection of *Edwadsiella tarda*). 18 colonies from the feces in FE-control and 21 colonies from the feces in FE-5% compost were isolated. The ratio of bacteria identified on the sequence of 16Sr RNA is shown in each group. *B.* shows the genus *Bacillus* containing a common synonym. *B.coagulans*, *B.thermoamylovorans*/*hisashii*, and *B.megaterium*, a common synonym and recently recategorized as *Weizmannia coagulans*, *Caldibacillus thermoamylovorans*/*hisashii*, and *Priestia megaterium*, respectively[26]. The marks shows as follows: #, p < 0.1;*, p < 0.05; **, p < 0.01

Extended fig.2

(a) The time series of body weights of chicks in the CC Test were shown. The number in parentheses shows the number of individual chicks. The averaged body weights in each week shown as a unit of gram. (b) shows the population of *Salmonella* in the cecum and ovary tract after the *Salmonella*-injected experiment in CC-test (Fig.S.3a). (c) shows the population of a genus of the bacterial community in the feces of chicks of CC-test (n=3 as group sample). (d) shows relative abundances of a detected genus of fecal microbiota before and after the compost extract in the CW test. (e) shows fecal IgA contents at 3 month after compost exposure in CW test. (n=7). (f) the integrated values of weekly death rates in the hen house (n=54488-55618 per each hen house as number of introduced hens) were compiled weekly in CCR-test. CCR-normal (Positive control) shows data of a hen house fed under normal conditions (around 20-25 °C as average temperature). CCR-control 1 (Negative control 1) and CCR-control 2 (Negative control 2) show data of two hen houses in summer (around 25-30 °C as average temperature). CCR-compost ex shows data of hen house with compost exposure in the same summer. (g) shows the scatter plot for the relationship between the averages of temperature and death rates of hen (per week) in the CCR-test (n=73-77 as data of each hen house). (h) shows the relative population of a genus of the fecal bacterial community of hen in the tested hen house in the CCR farm (n=5; >1% as bacterial population). The data in (a), (b), and (e) were shown as Means ± SE. The mark of asterisks * shows p < 0.05.

Extended fig.3

(a) The bacterial population of a significant genus in the feces of piglets in the PK and PF swine farm (Fig.S5a). (b) shows the death rates of growing pigs and the times of administrated antibiotic drugs in the PI swine farm (Fig.S5a) (c) the death rates of growing pigs (Male and Female) and (d) their cost of drugs per individual swine house in PM swine farm (Fig.S5a) are shown. The data of 106 swine houses (around 400 individual pigs per swine house; the sum of individual pigs is 462,336) is calculated. The data of three houses (approximately 400 individual pigs per house) as growing swine houses after the exposure of thermophile *C. hisashii* were marked with red dots. Female A and Female B were discriminated as the groups fed with different feed, respectively. (e) In PM swine farm, fecal IgA contents of piglets with exposure of thermophile *C. hisashii* were shown (p=0.11). (f) shows the fecal contents of acetate, propionate, butyrate, and lactate in the PM farm. (g) shows a scatter plot for the relationship between IgA, SCFAs, and lactate in the PM farm. SCFAs is the sum of acetate, propionate, and butyrate. (h) phyla as the fecal bacterial population in the PM farm were shown. (i) shows the scatter plot for the relation between Firmicutes, Actinobacteria (orange circle), and Bacteroidetes (blue circle) in the PM farm.

Extended fig.4

The photograph of feces from a calve with (a) diarrhea and (b) their recovered feces after the thermophile *C.* administration in KC farm, respectively. (c) shows fecal IgA contents in their calves (a) and (b) and calves with standard conditions. (d) shows the contents of SCFAs in the (a) and (b). (e) shows the scatter plot for the relation between IgA and SCFAs in the KC farm. (f) shows the scatter plots for the relationship between SCFA contents and $H_2O$ in the feces. (g) shows fecal bacterial population before the administration (Before) and thereafter (After). All the calves before the administration were with diarrhea. (h) shows the bacterial population of genus *Fusobacteirum* before the administration (Before) and thereafter (After). (i) shows the bacterial population of close related species, which showed significant differences between before the administration (Before) and thereafter (After) (p<0.05).

**Supplementary table legends**

Table S1.

A model and its fit indices inconcluded with amino acids were shown in the No.1 column. The No.2-No.4 shows the inferior models and their fit indices.

Table S2.

A model and its fit indices in Figs.2c and 2d were shown in the No.1 and No.2 column. The No.2-No.4 shows the inferior models and their fit indices.

**Supplementary figure legends**

Fig.S1

(a) α-diversity, (b) β-diversity, (c) population of phylum, (d) population of genus of the fecal microbiota of fish, chicken, pig, and cattle.

Fig.S2

(a) shows the scheme of the fish tests to evaluate the effects of the compost against *Edwadsiella tarda* in Shizuoka (FS-Test) and Ehime (FE-Test), respectively. (b) shows recovery time of the fish after anesthesia. (c) shows the physiological conditions of red breams in FE-test. (d) shows the contents of fecal organic acids (SCFA, lactate and succinate) in fish without the compost exposure, FE-control, with 1% compost, and with 5% compost in FE-test. (e ) and (f) shows β-diversity and their $R^2$ and p values of unweighted unifrac distances and the weighted unifrac distances, respectively. (g) and (h) shows the population of phylum and genus of the fecal bacterial community of fish tested in FE-Test. and significantly changed the top three substances (b)-(d) in fish of FE-control and FE- 5% compost in Ehime. #, $p <0.1$; *, $p < 0.05$; **, $p < 0.01$.

Fig.S3

(a) shows the scheme of the chicken (hen) tests to evaluate the effects of the compost. (b) α-diversity, (c) β-diversity, (d) population of phylum, and genus of the fecal microbiota in CC-Test were shown. (e) α-diversity and (f) β-diversity of the fecal microbiota in CW-Test were shown.

Fig.S4

(a) shows the data of two hen houses increased death rate (Control 1 and 2) in the summer and ones of six hen houses administrated the compost extract (Test 1-6) in the same duration. (b) shows OPG values in the tested hen house (n=4-5/month per group). *, $p<0.05$; #, $p<0.2$ (c) the contents of organic acids (SCFA and lactate) in feces of hen in the summer before and after the compost extract test in the CCR farms were shown. ***, $p<0.001$; **, $p<0.01$;*, $p<0.05$; #, $p<0.1$; $, $p<0.2$. (d) α-diversity, (e) β-diversity, and (f) population of species in the fecal bacterial community of chicken in the CCR farm were shown ($p<0.05$). The feces of chickens over 6 months with and without the administration of the compost extract in the same henhouse were detected.

Fig.S5

(a) shows the scheme of the pig tests to evaluate the effects of the compost and thermophile. (b) α-diversity, (c) β-diversity, (d) their $R^2$ and p values of unweighted unifrac distances and the weighted unifrac distances, and (e) population of the phylum of the fecal microbiota in the PK TEST and PF TEST were shown, respectively.

Fig.S6

(a) IgA contents and (b) organic acids (SCFA and lactate) in the feces of pigs in the swine farm of PI Test were shown. (c) α-diversity, (d) β-diversity, (e) population of phylum, and (f) population of the genus of the fecal bacterial community of pigs in the swine farm were shown. (g) Fecal IgA contents in the swine farm of PM Test were shown Sow No. shows individual number. (n=5 ; the number of the piglets produced from an individual sow). The mark of asterisks * shows $p < 0.05$, and asterisks *** shows $p < 0.001$.

Fig.S7

(a) IgA contents and (b) organic acids (SCFA and lactate) in the feces of pigs in the swine farm of PM Test were shown. (c) α-diversity, (d) β-diversity, (e) population of phylum, and (f) population of genus of the fecal bacterial community of pigs in the swine farm were shown.

Fig.S8

(a) shows the scheme of the cattle tests to evaluate the effects of thermophile *C. hisashii* in KC Test. (b) shows the fecundity conditions of calves in KC Test. (c) shows the difference of $H_2O$ ratio (%) in normal feces (n=8), in feces of calves before (n=4) and after (n=4) *thermophile C.* administration. (d) α-diversity, (e) β-diversity, (f) population of the phylum of the fecal microbiota in the cattle farm were shown. (g) shows individual positive pathogenic genes of *Escherichia coli*.

Fig.S9

Heatmaps showing the relative abundance of metabolites in feces of (a) fish and (b) chicken ($p <0.3$) (#, $p <0.1$; *, $p<0.5$; **, $p<0.01$).

Fig.S10

Heatmaps showing the relative abundance of metabolites in feces of (c) pig and (b) cattle ($p <0.3$)(#, $p <0.1$; *, $p<0.5$; **, $p<0.01$).

Fig.S11

Correlation heatmap of the fecal metabolites of all animals, The compounds were selected if there was any significant difference ($p <0.1$) in one species and more of the animals.

Fig.S12

(a) overview of the pathway analyses of the molecular metabolic mechanism of the gut environments altered by oral administration of compost and/or thermophile *C. hisashii* to all animals. (b) overview of the pathway analyses of the molecular metabolic mechanism of the gut environments altered by oral administration of compost and/or thermophile *C. hisashii* to all animals. The mark of F, Ch, Pc, Pt, and Ca shows the values of fish, chicken, pig, and cattle of test groups, respectively. Pc and Pt show pigs administrated with compost and with the thermophile, respectively.

Fig.S13

(a) Interactive systemic networks represented by association analysis using whole phyla, genera (> 0.1% as detection rate), and all the metabolites detected in all the animals (n=4). The intensely interactive components, targets associated directly by the compost- and thermophile-administrated condition as a source data (> lift value 1.2), were selected, and the representative networks with the components were shown by Gephi. Two colors discriminate between modularity classes. The difference in the color of nodes indicates the strength of degree, whose value sums up the weights of the adjacent edges for each node. (b) interactive systemic networks represented by association analysis using whole phyla, genera (> 0.1% as detection rate), and all the metabolites detected in all the animals (n=4 per each group). The intensely interactive components, targets associated directly by the compost-administrated condition as a source data (> lift value 1.2), were selected, and the representative networks with the components were shown by Gephi. Modularity classes are discriminated by three colors. The difference in the color of nodes indicates the strength of degree, whose value sums up the weights of the adjacent edges for each node. (c) Relative abundance of selected fecal bacteria was assessed for each animal.

Fig. S14

Relative ratio of fecal components selected by association analysis shows in of fish, chicken, pig, and cattle. Blue: control; Red: compost group; Yelow: *Caldibacillus hisashii* administered group.

Fig. S15

Response to Environmental $\varepsilon$. Dependencies to (a) bacterial and (b) metabolite stage were plotted. The right and left in X axis was plus and minus effects, respectively.

Fig. S16

The interaction networks of optimal components based on SEM were visualized as the significant relationships in the extended pairwise maximum entropy model: (a) Lactobacillus; (b) AIB; (c) Butyrate. The green and violet lines show positive and negative effects between the components, respectively..

Fig. S17

When treated with (a) vegetative cell and spore for long term (151 days) and (b) short term (35 days) of *Caldibacillus hisashii*, metabolome analysis data are shown ($p<0.05$). Double asterisk indicates $p<0.01$.

Fig. S18

Structure prediction of papain like-protease of *Caldibacillus hisashii* calculated by AlphaFold2. It contains stop codons on the genome sequence, and therefore, the stop codon was removed and predicted as a process, taking into account and discounting the splicing. Based on the (a) Predicted Aligned Error (PAE) and (b) predicted local distance difference test (pLDDT), five prediction data (rank_1 - rank_5) were calculated. (c) The predicted three-dimensional structure is visualized with an interpretation based on the calculation results of pLDDT.

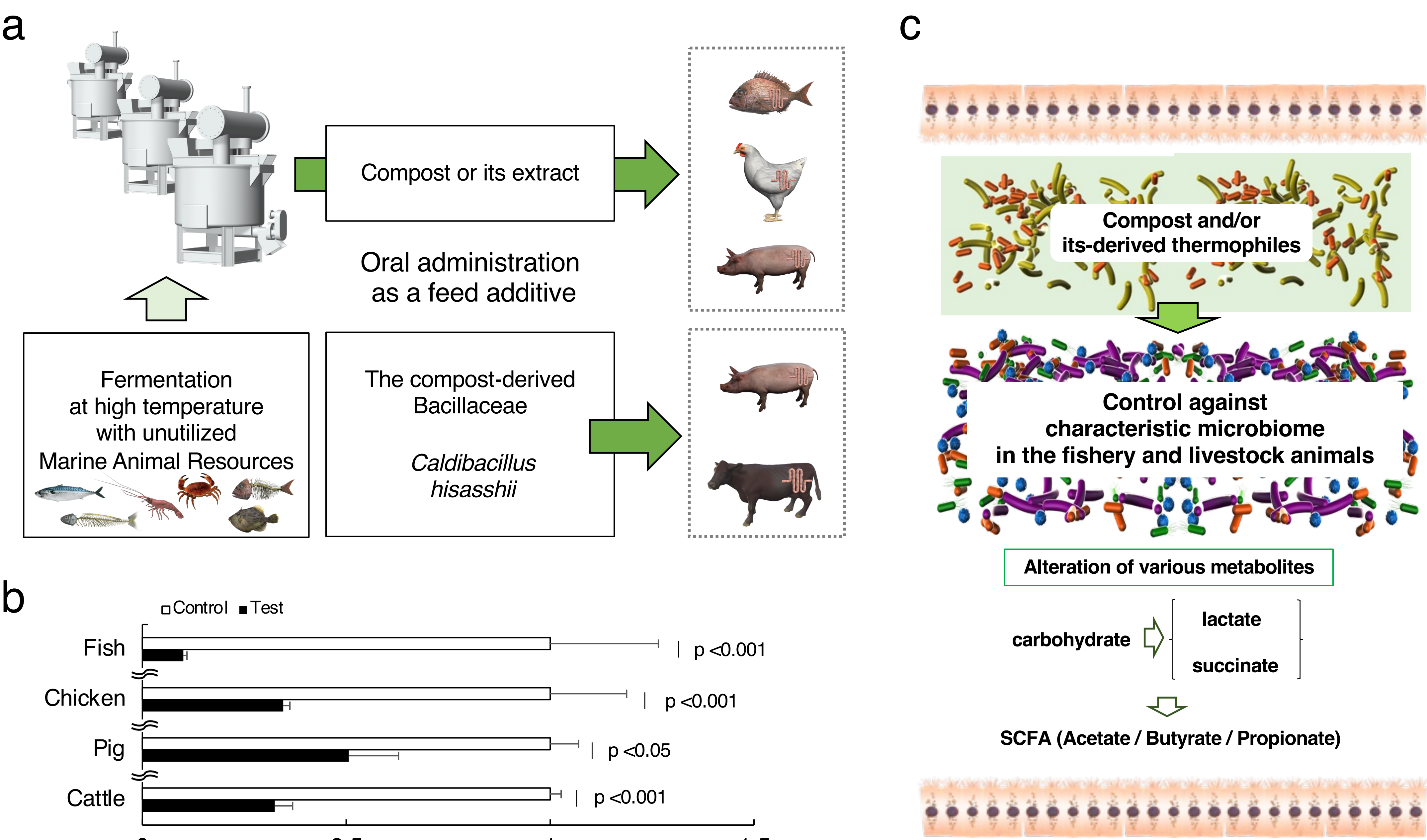

Fig.1

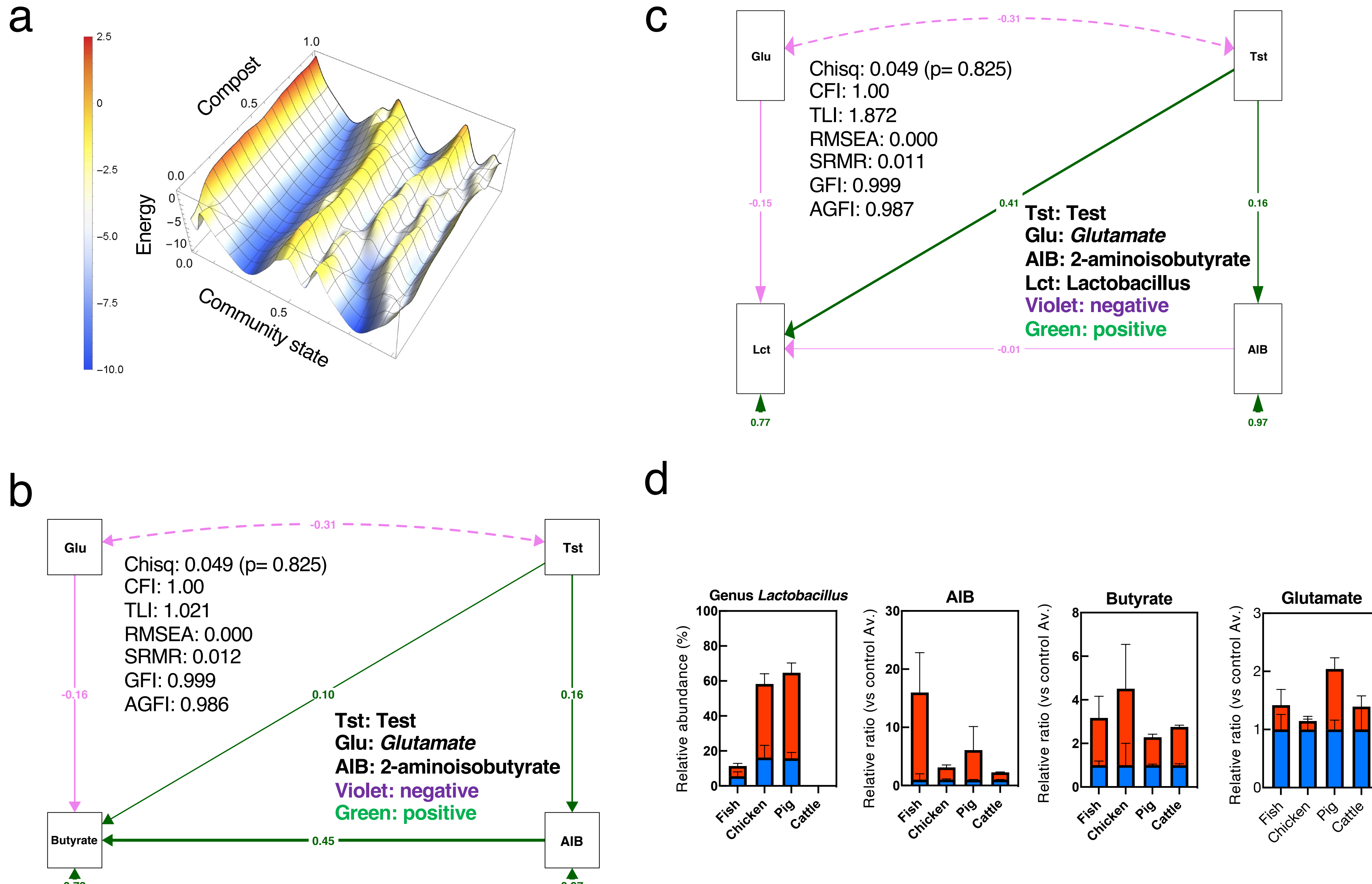

Fig.2

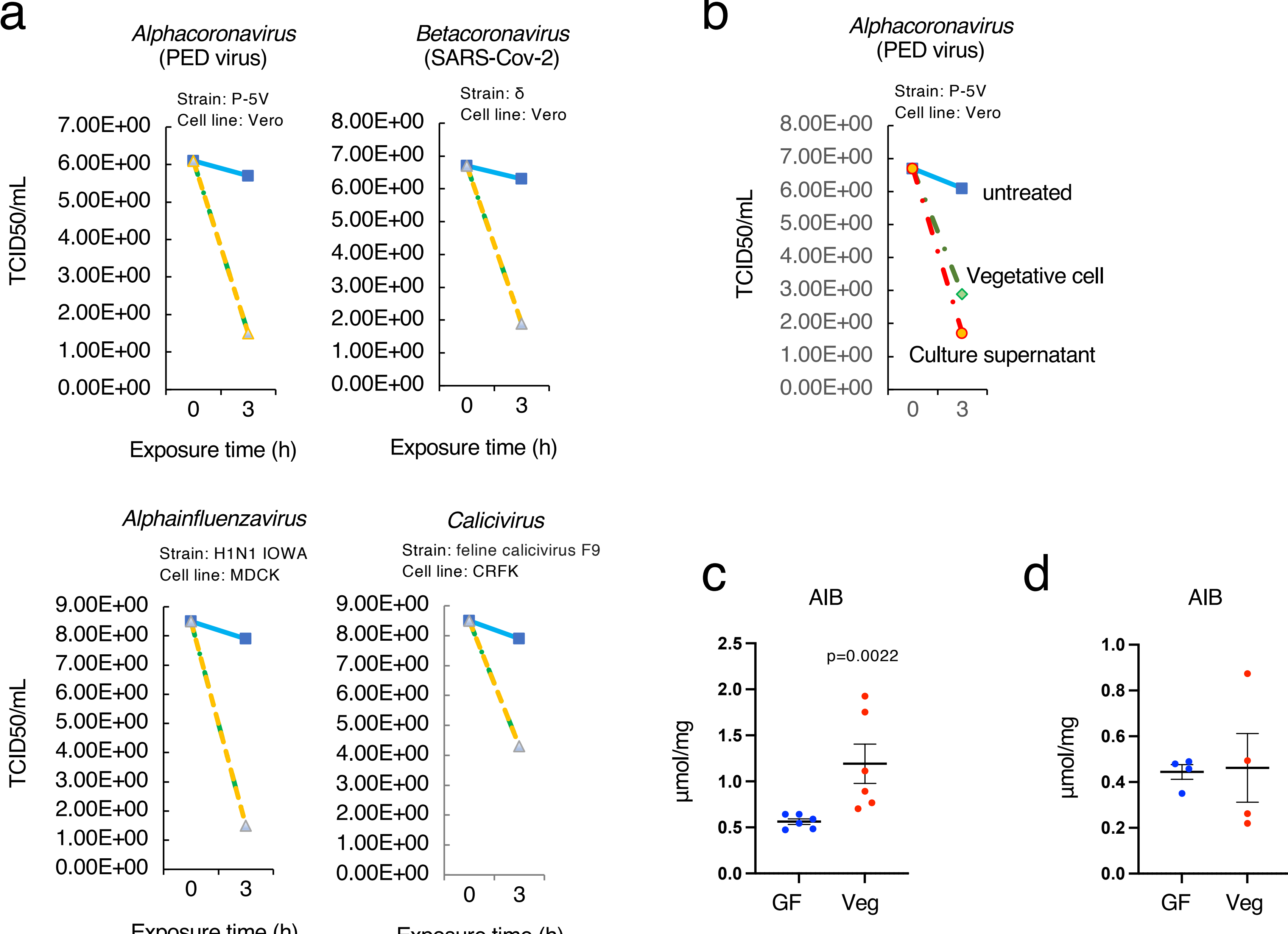

Fig.3

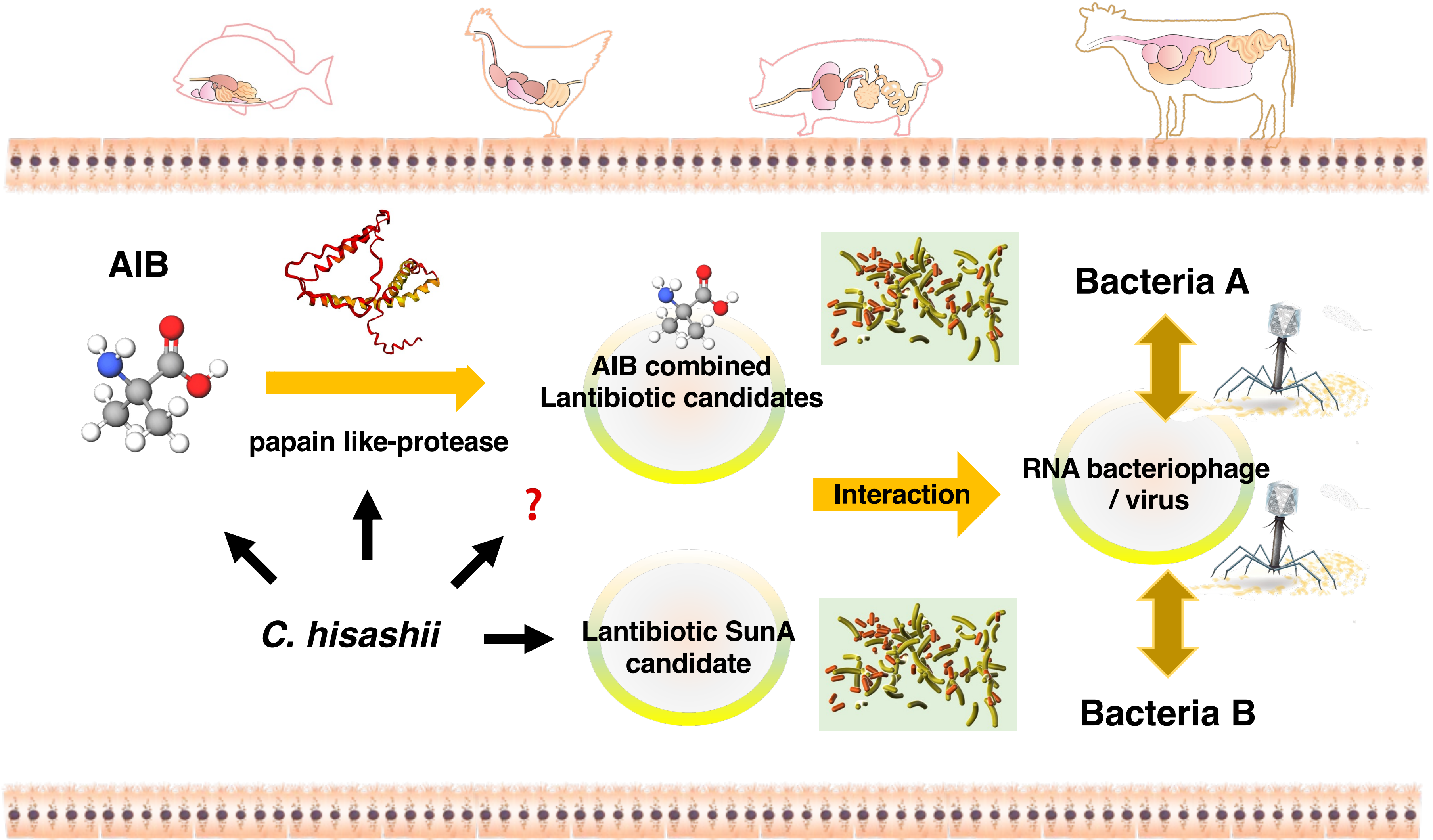

Fig.4

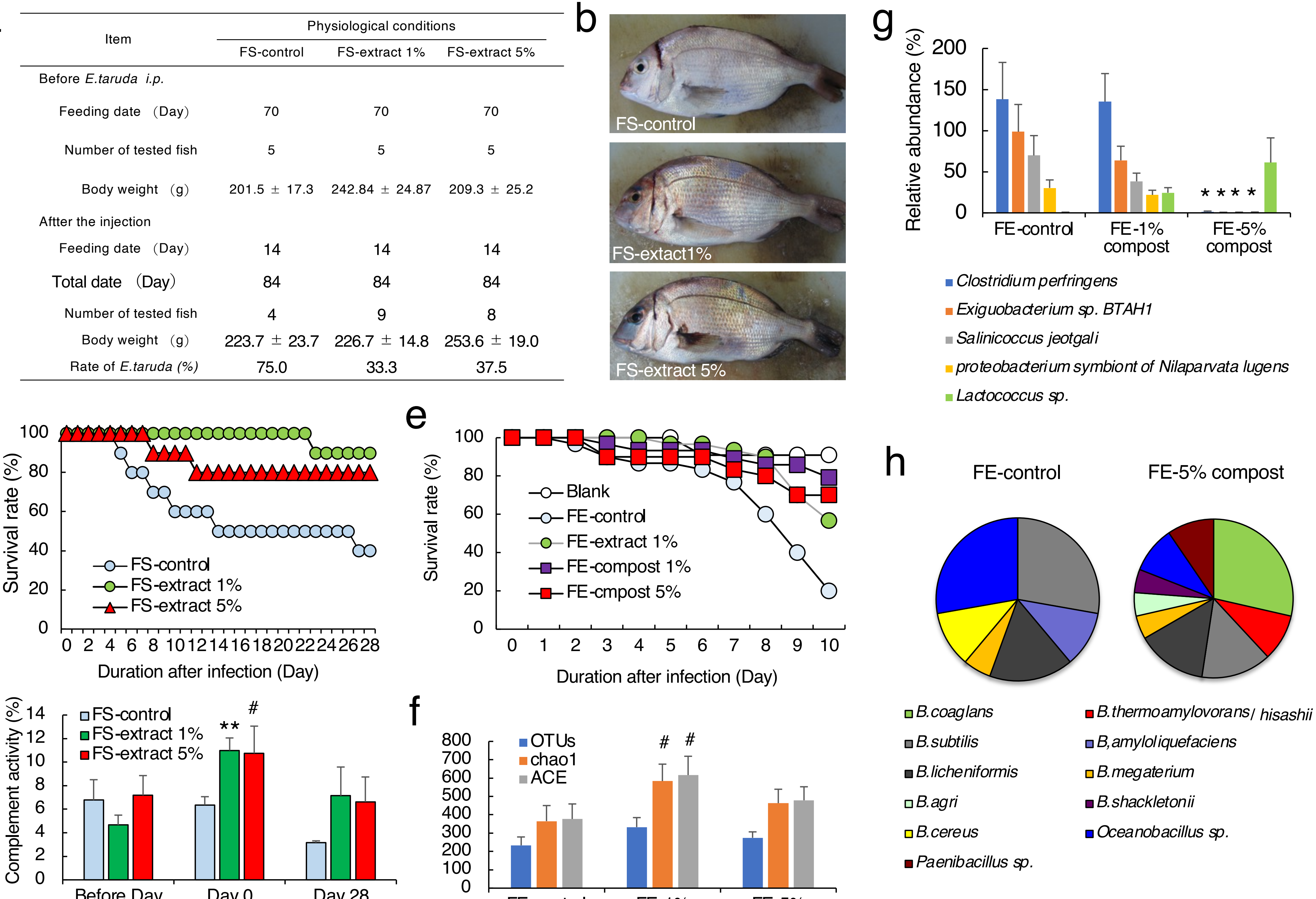

| Item | Physiological conditions | | |
| --- | --- | --- | --- |
| | FS-control | FS-extract 1% | FS-extract 5% |
| Before *E.taruda* *i.p.* | | | |
| Feeding date (Day) | 70 | 70 | 70 |
| Number of tested fish | 5 | 5 | 5 |
| Body weight (g) | 201.5 ± 17.3 | 242.84 ± 24.87 | 209.3 ± 25.2 |
| Aftter the injection | | | |
| Feeding date (Day) | 14 | 14 | 14 |
| Total date (Day) | 84 | 84 | 84 |
| Number of tested fish | 4 | 9 | 8 |
| Body weight (g) | 223.7 ± 23.7 | 226.7 ± 14.8 | 253.6 ± 19.0 |
| Rate of *E.taruda* (%) | 75.0 | 33.3 | 37.5 |

Extended fig.1

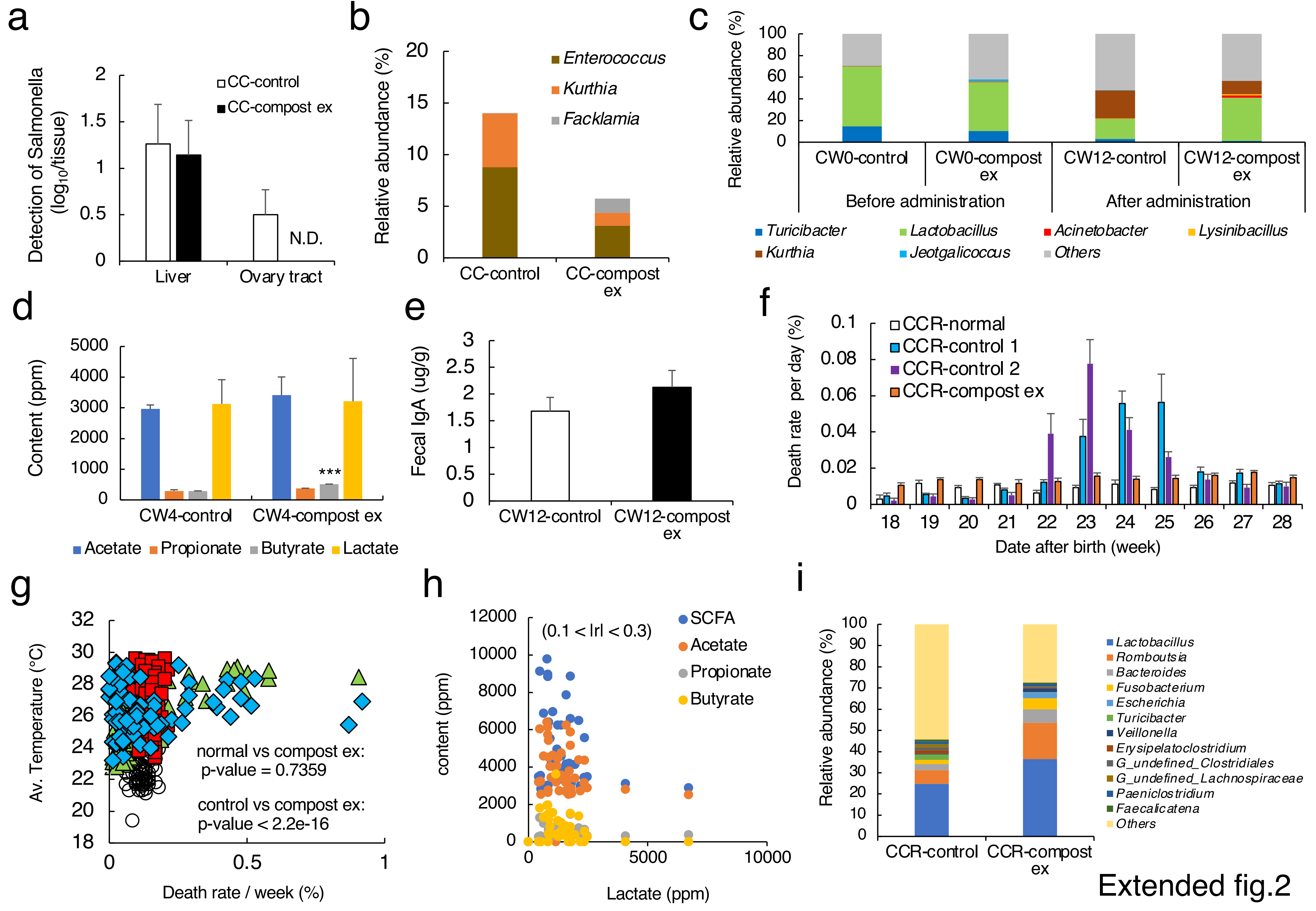

Extended fig.2

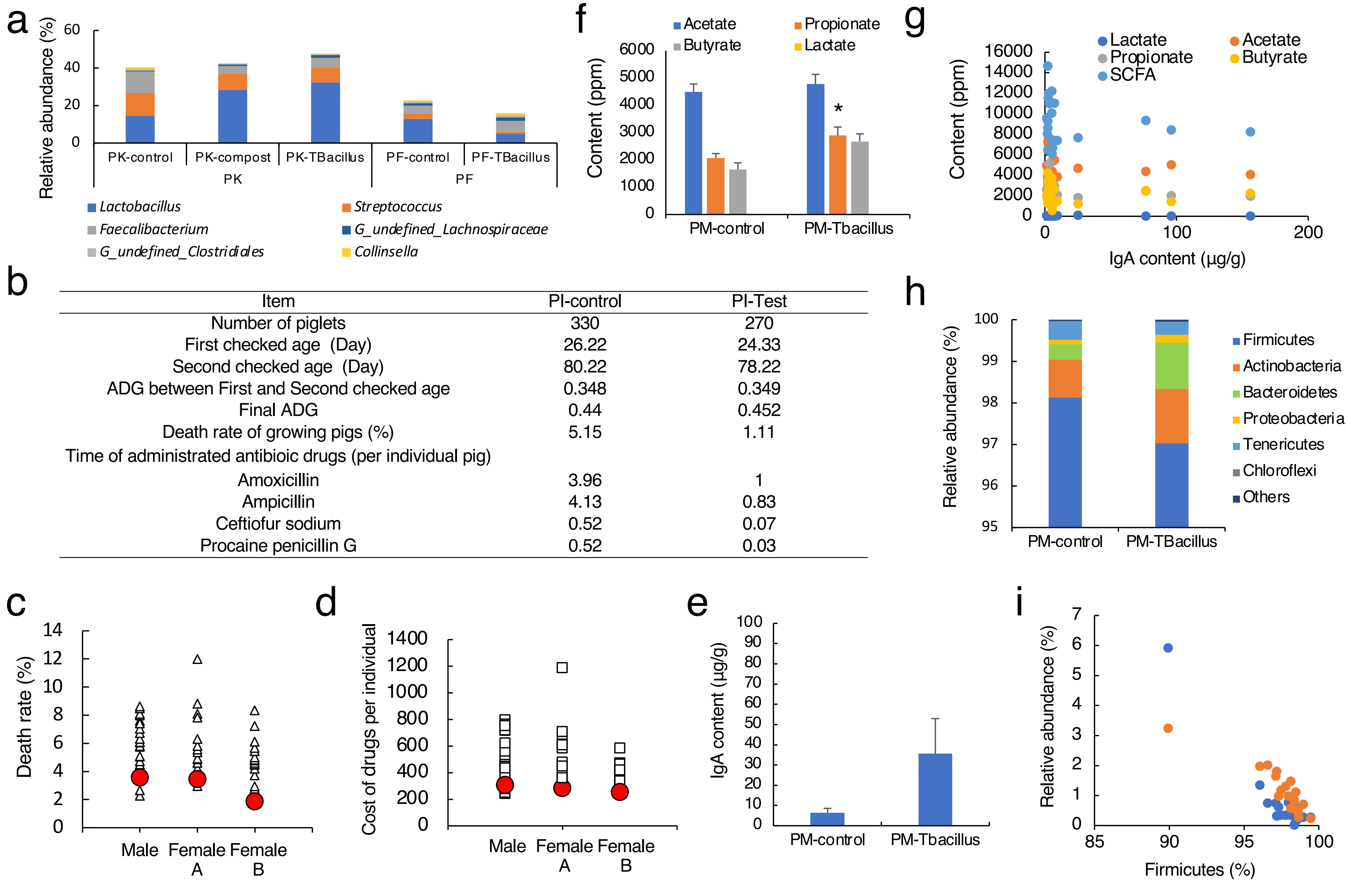

| Item | PI-control | PI-Test |
|---|---|---|
| Number of piglets | 330 | 270 |
| First checked age (Day) | 26.22 | 24.33 |
| Second checked age (Day) | 80.22 | 78.22 |
| ADG between First and Second checked age | 0.348 | 0.349 |
| Final ADG | 0.44 | 0.452 |
| Death rate of growing pigs (%) | 5.15 | 1.11 |
| Time of administrated antibioic drugs (per individual pig) | | |
| Amoxicillin | 3.96 | 1 |
| Ampicillin | 4.13 | 0.83 |
| Ceftiofur sodium | 0.52 | 0.07 |
| Procaine penicillin G | 0.52 | 0.03 |

Extended fig.3

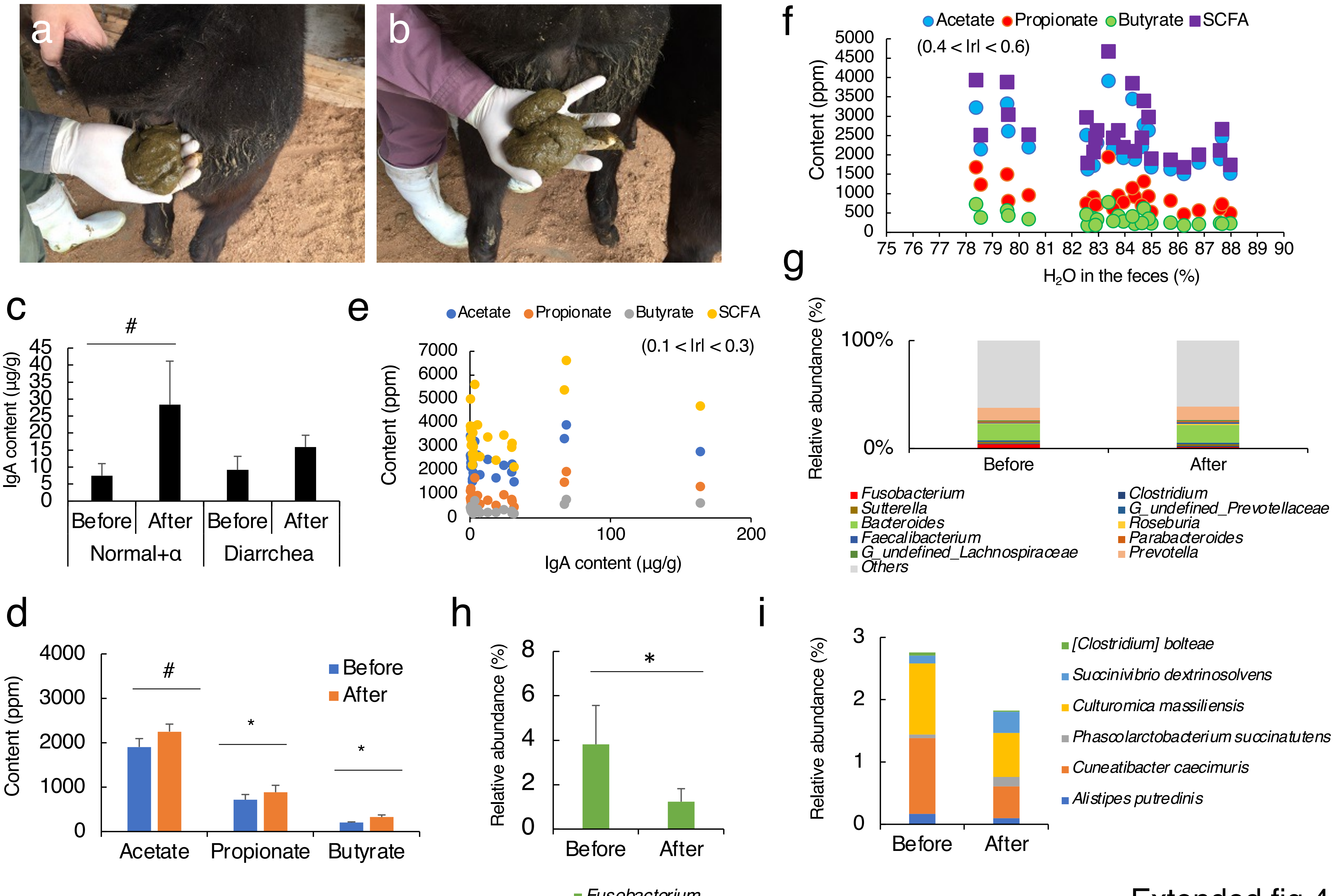

Extended fig.4

# Table S1

| Category | Model | Fit indices | | |
|---|---|---|---|---|
| No.1 | Aminoacid_content ~ Tryptamine + Tyrosine + Glutamate + Tryptophan | chisq 25.807 | df 27.000 | pvalue 0.529 |
| | Lactobacillus ~ Test | cfi 1.000 | tli 1.004 | rfi 0.918 |
| | Hydroxyisobutyrate + Aminovalerate ~ Test | nfi 0.937 | srmr 0.129 | AIC 1703.157 |
| | AIB ~ Test | rmsea 0.000 | gfi 0.999 | agfi 0.998 |
| No.2 | Aminoacid_content ~ Tryptamine + Tyrosine + Glutamate | chisq 24.779 | df 23.000 | pvalue 0.362 |
| | Lactobacillus ~ Test | cfi 0.993 | tli 0.991 | rfi 0.891 |
| | Hydroxyisobutyrate + Aminovalerate ~ Test | nfi 0.917 | srmr 0.126 | AIC 1701.829 |
| | AIB ~ Test | rmsea 0.046 | gfi 0.999 | agfi 0.998 |
| No.3 | Aminoacid_content ~ Tryptamine + Tyrosine + Glutamate + Tryptophan | chisq 24.135 | df 19.000 | pvalue 0.191 |
| | Lactobacillus ~ Test | cfi 0.987 | tli 0.982 | rfi 0.919 |
| | Hydroxyisobutyrate + Aminovalerate ~ Test | nfi 0.941 | srmr 0.139 | AIC 1311.559 |
| | | rmsea 0.087 | gfi 0.999 | agfi 0.998 |
| No.4 | Aminoacid_content ~ Tyrosine + Glutamate | chisq 23.293 | df 19.000 | pvalue 0.225 |
| | Lactobacillus ~ Test | cfi 0.978 | tli 0.971 | rfi 0.863 |
| | Hydroxyisobutyrate + Aminovalerate ~ Test | nfi 0.896 | srmr 0.139 | AIC -1703.832 |
| | | rmsea 0.079 | gfi 0.999 | agfi 0.998 |

# Table S2

| Category | Model | Fit indices | | |
|---|---|---|---|---|
| No.1 | Lactobacillus~ AIB + Glutamate | chisq 0.049 | df 1.000 | pvalue 0.825 |
| | Lactobacillus + AIB ~ Test | cfi 1.000 | tli 1.872 | rfi 0.976 |
| | AIB ~ Test | nfi 0.995 | srmr 0.011 | AIC 713.074 |
| | | rmsea 0.000 | gfi 0.999 | agfi 0.987 |
| | lavaan 0.6-12 ended normally after 1 iterations<br>standard errors number of successful bootstrap draws : 1000 | | | |
| No.2 | Butyrate ~ Glutamate + AIB | chisq 0.049 | df 1.000 | pvalue 0.825 |
| | Butyrate + AIB ~ Test | cfi 1.000 | tli 1.021 | rfi 0.999 |
| | | nfi 1.000 | srmr 0.012 | AIC 1182.175 |
| | | rmsea 0.000 | gfi 0.999 | agfi 0.986 |
| | lavaan 0.6-12 ended normally after 1 iterations<br>standard errors number of successful bootstrap draws : 1000 | | | |
| No.3 | Lactobacillus ~ Tryptamine | chisq 0.025 | df 1.000 | pvalue 0.875 |
| | Lactobacillus + AIB + Triptamine ~ Test | cfi 1.000 | tli 2.708 | rfi 0.984 |
| | AIB ~ Tryptamine | nfi 0.997 | srmr 0.007 | AIC 754.648 |
| | lavaan 0.6-12 ended normally after 1 iterations<br>standard errors number of successful bootstrap draws : 1000 | rmsea 0.000 | gfi 1.000 | agfi 0.996 |
| No.4 | Lactobacillus ~ Tryptamine + Glutamic_acid | chisq 0.266 | df 3 | pvalue 0.966 |
| | Lactobacillus + AIB ~ Test | cfi 1.000 | tli 2.662 | rfi 0.943 |
| | | nfi 0.975 | srmr 0.021 | AIC 712.904 |
| | lavaan 0.6-12 ended normally after 1 iterations<br>standard errors number of successful bootstrap draws : 1000 | rmsea 0.000 | gfi 0.993 | agfi 0.965 |

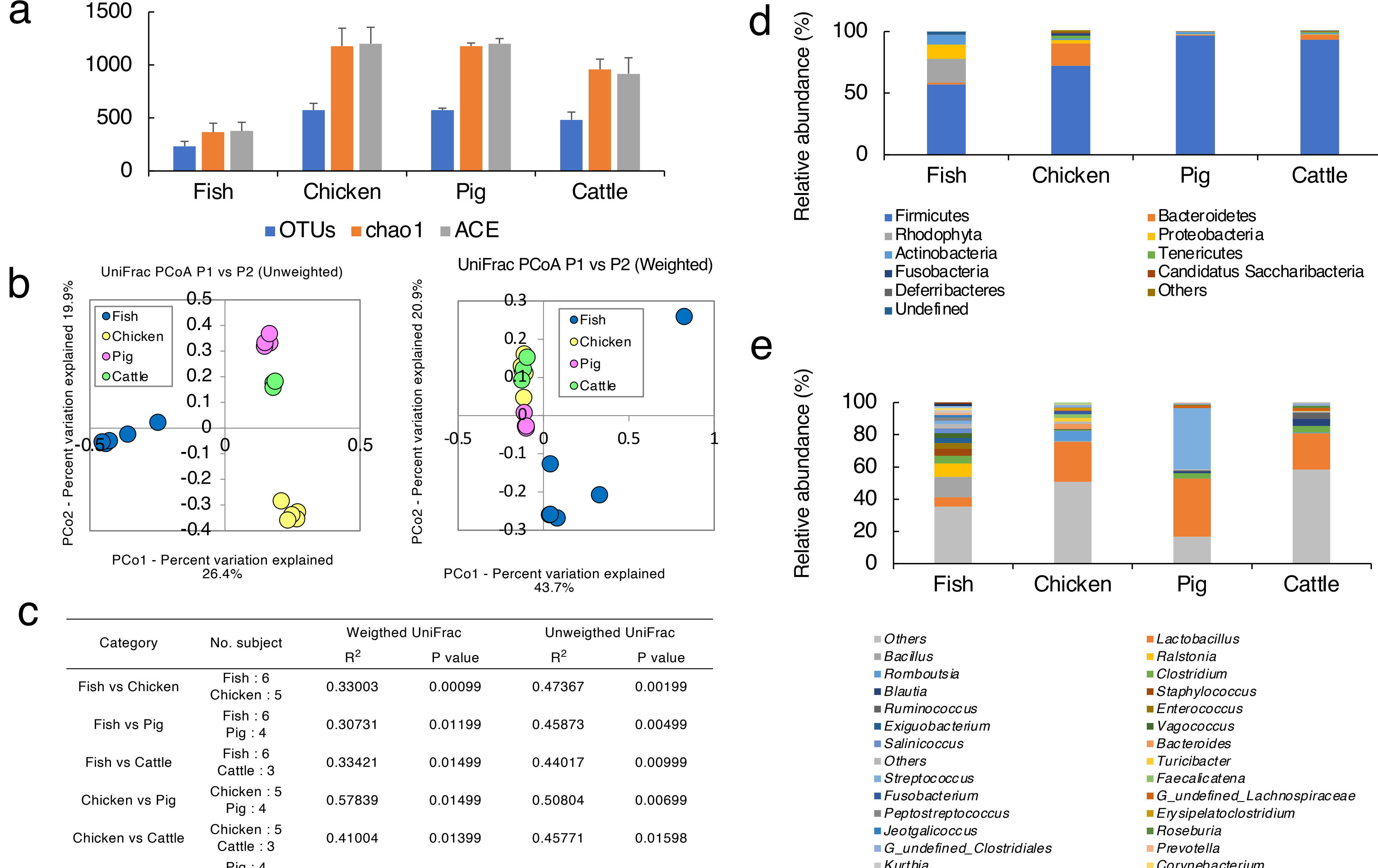

| Category | No. subject | Weigthed UniFrac | | Unweigthed UniFrac | |
|---|---|---|---|---|---|
| | | $R^2$ | P value | $R^2$ | P value |
| Fish vs Chicken | Fish : 6<br>Chicken : 5 | 0.33003 | 0.00099 | 0.47367 | 0.00199 |
| Fish vs Pig | Fish : 6<br>Pig : 4 | 0.30731 | 0.01199 | 0.45873 | 0.00499 |
| Fish vs Cattle | Fish : 6<br>Cattle : 3 | 0.33421 | 0.01499 | 0.44017 | 0.00999 |
| Chicken vs Pig | Chicken : 5<br>Pig : 4 | 0.57839 | 0.01499 | 0.50804 | 0.00699 |
| Chicken vs Cattle | Chicken : 5<br>Cattle : 3 | 0.41004 | 0.01399 | 0.45771 | 0.01598 |
| Pig vs Cattle | Pig : 4<br>Cattle : 3 | 0.58319 | 0.03197 | 0.44211 | 0.02797 |

Fig.S1

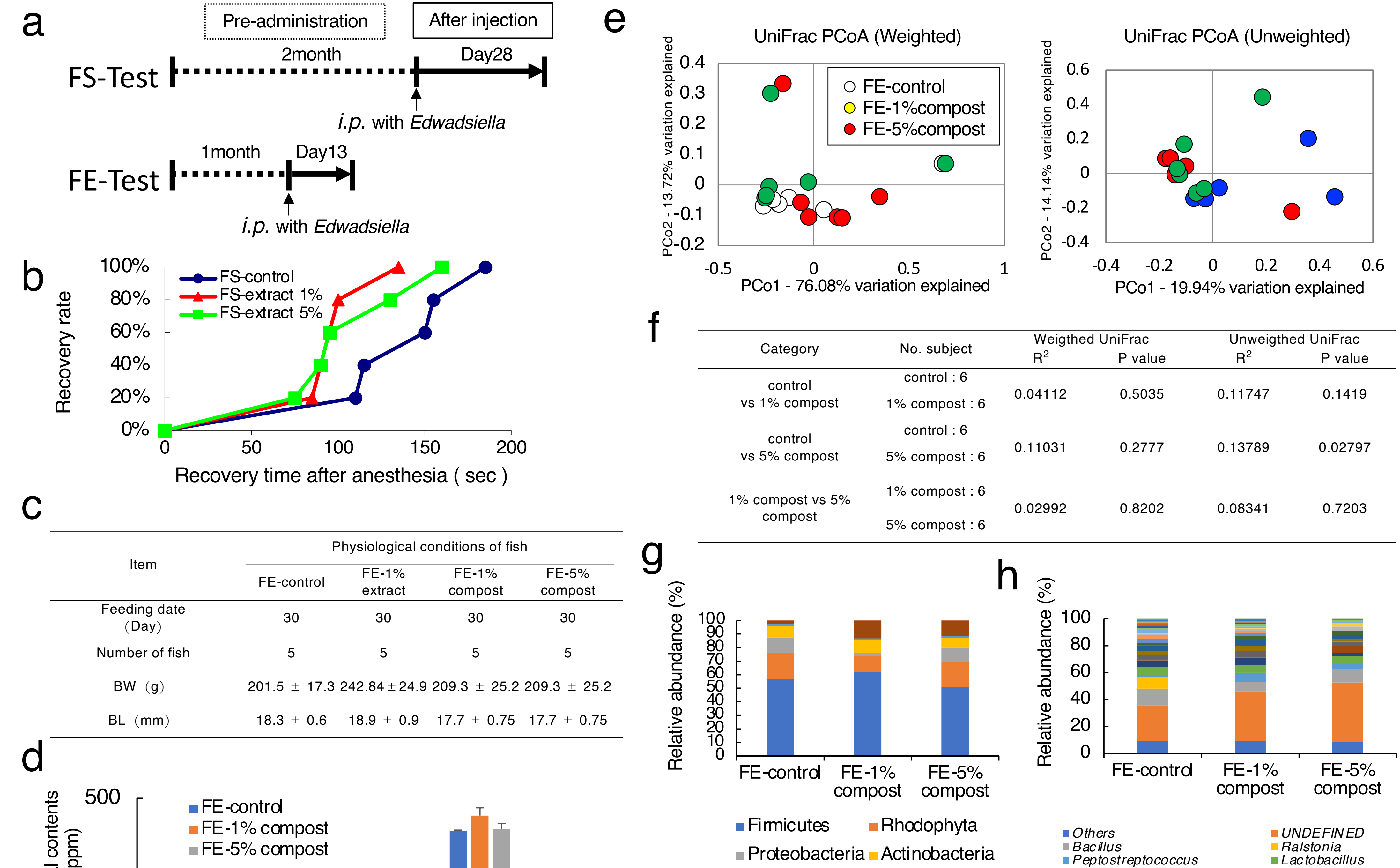

c

| Item | Physiological conditions of fish | | | |
|---|---|---|---|---|
| | FE-control | FE-1% extract | FE-1% compost | FE-5% compost |
| Feeding date (Day) | 30 | 30 | 30 | 30 |
| Number of fish | 5 | 5 | 5 | 5 |
| BW (g) | 201.5 ± 17.3 | 242.84 ± 24.9 | 209.3 ± 25.2 | 209.3 ± 25.2 |
| BL (mm) | 18.3 ± 0.6 | 18.9 ± 0.9 | 17.7 ± 0.75 | 17.7 ± 0.75 |

f

| Category | No. subject | Weigthed UniFrac | | Unweigthed UniFrac | |
|---|---|---|---|---|---|
| | | $R^2$ | P value | $R^2$ | P value |
| control vs 1% compost | control : 6<br>1% compost : 6 | 0.04112 | 0.5035 | 0.11747 | 0.1419 |
| control vs 5% compost | control : 6<br>5% compost : 6 | 0.11031 | 0.2777 | 0.13789 | 0.02797 |
| 1% compost vs 5% compost | 1% compost : 6<br>5% compost : 6 | 0.02992 | 0.8202 | 0.08341 | 0.7203 |

Fig.S2

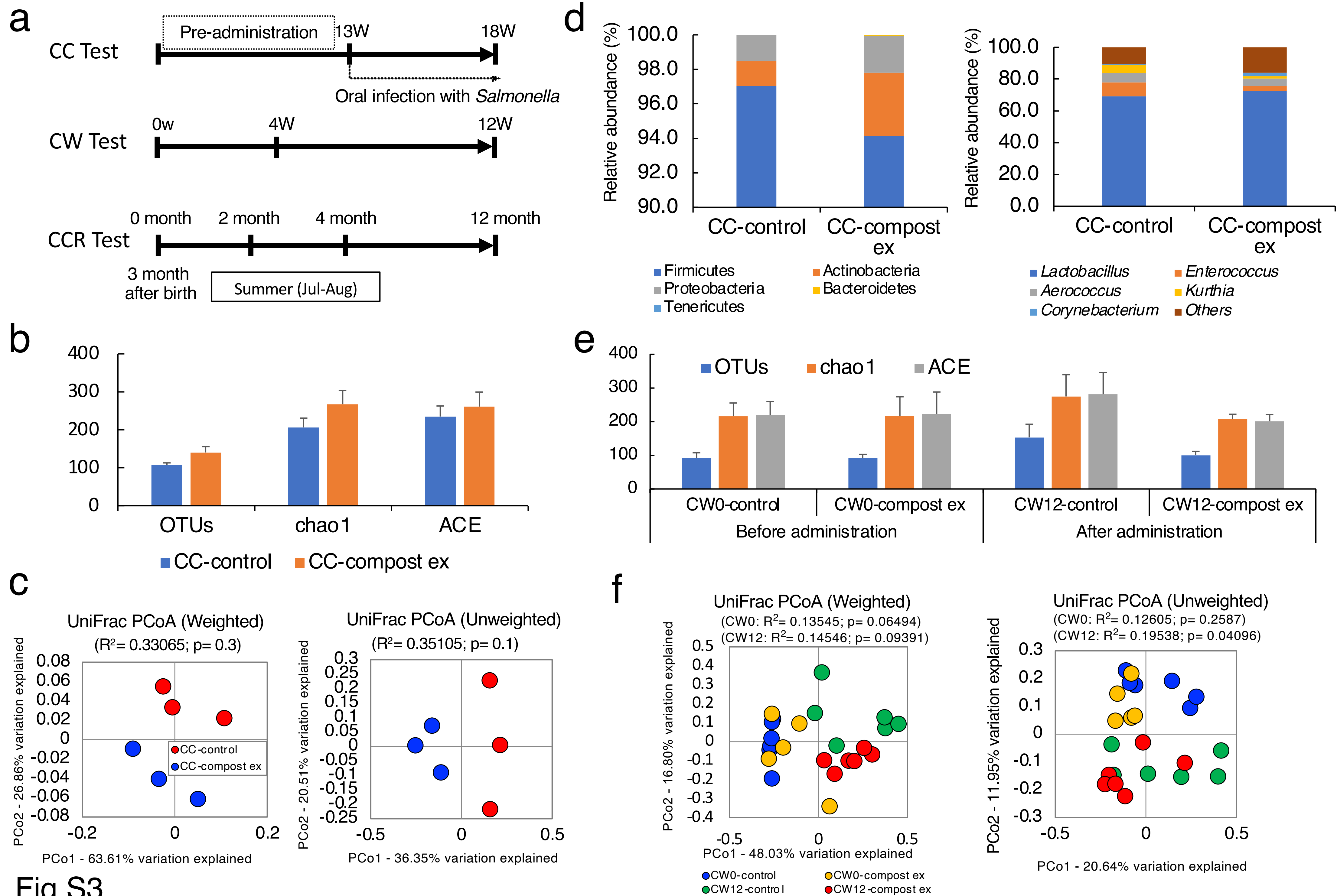

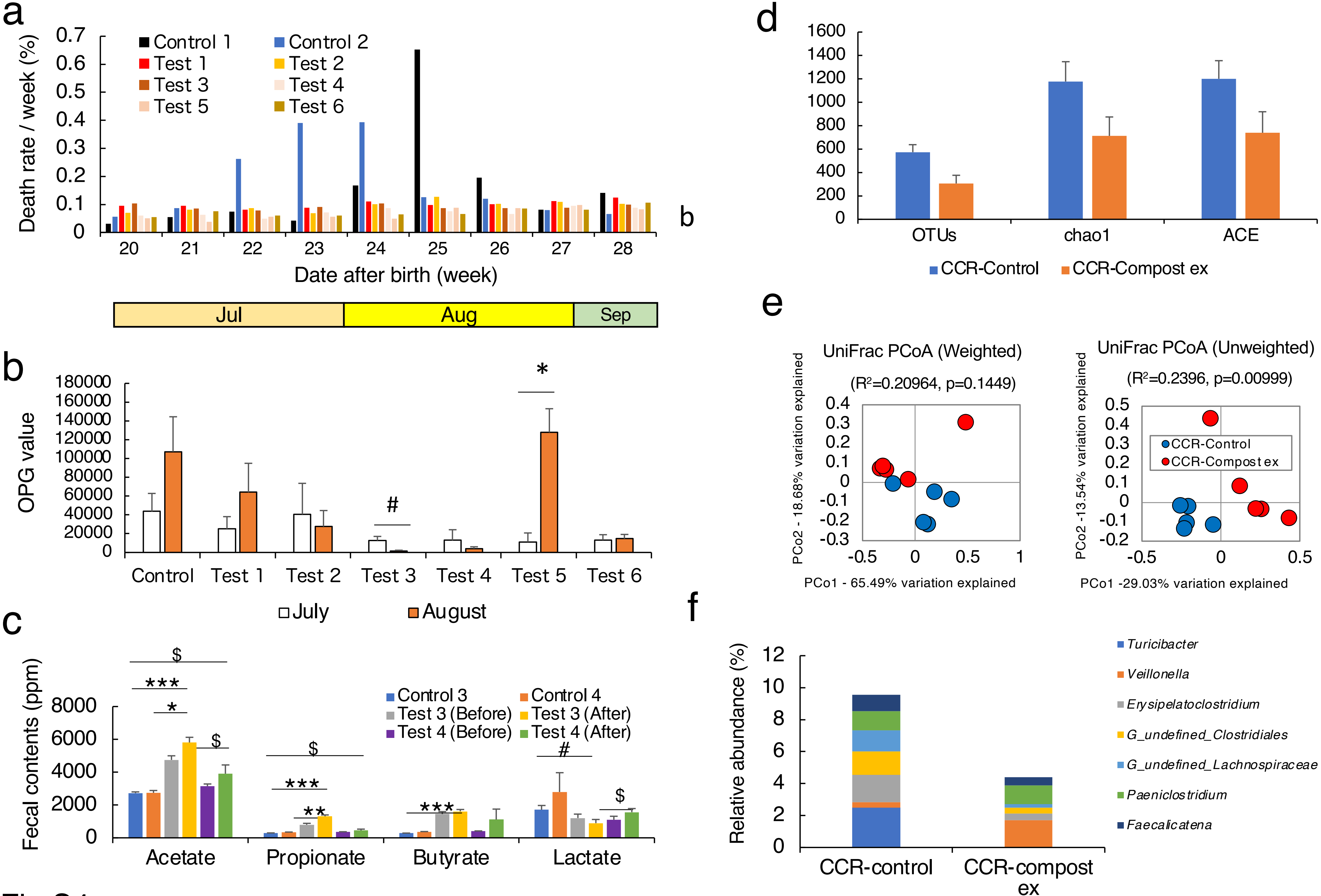

Fig.S4

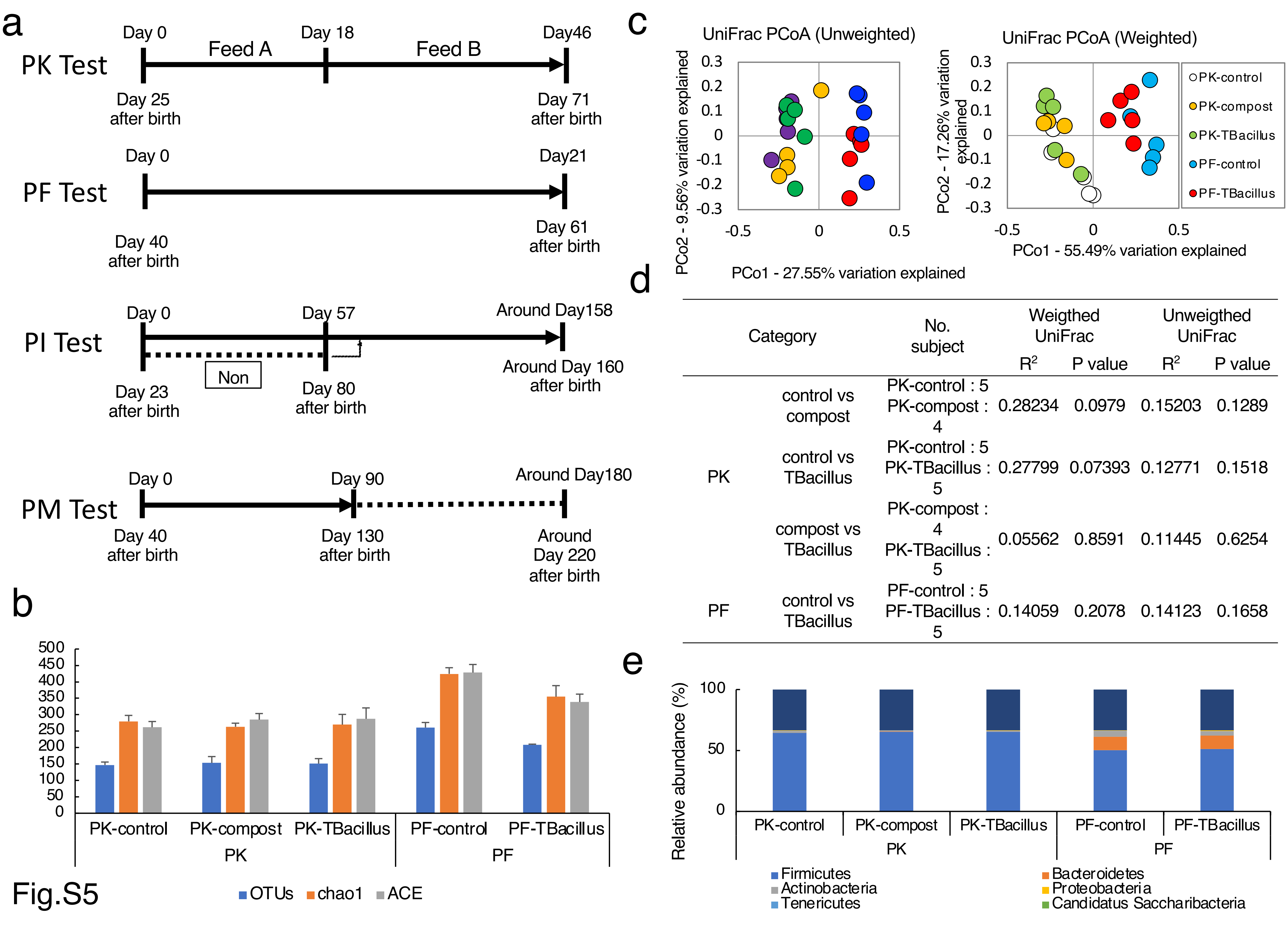

| Category | | No. subject | Weigthed UniFrac | | Unweigthed UniFrac | |
|---|---|---|---|---|---|---|
| | | | $R^2$ | P value | $R^2$ | P value |
| PK | control vs compost | PK-control : 5<br>PK-compost : 4 | 0.28234 | 0.0979 | 0.15203 | 0.1289 |
| | control vs TBacillus | PK-control : 5<br>PK-TBacillus : 5 | 0.27799 | 0.07393 | 0.12771 | 0.1518 |
| | compost vs TBacillus | PK-compost : 4<br>PK-TBacillus : 5 | 0.05562 | 0.8591 | 0.11445 | 0.6254 |
| PF | control vs TBacillus | PF-control : 5<br>PF-TBacillus : 5 | 0.14059 | 0.2078 | 0.14123 | 0.1658 |

Fig.S5

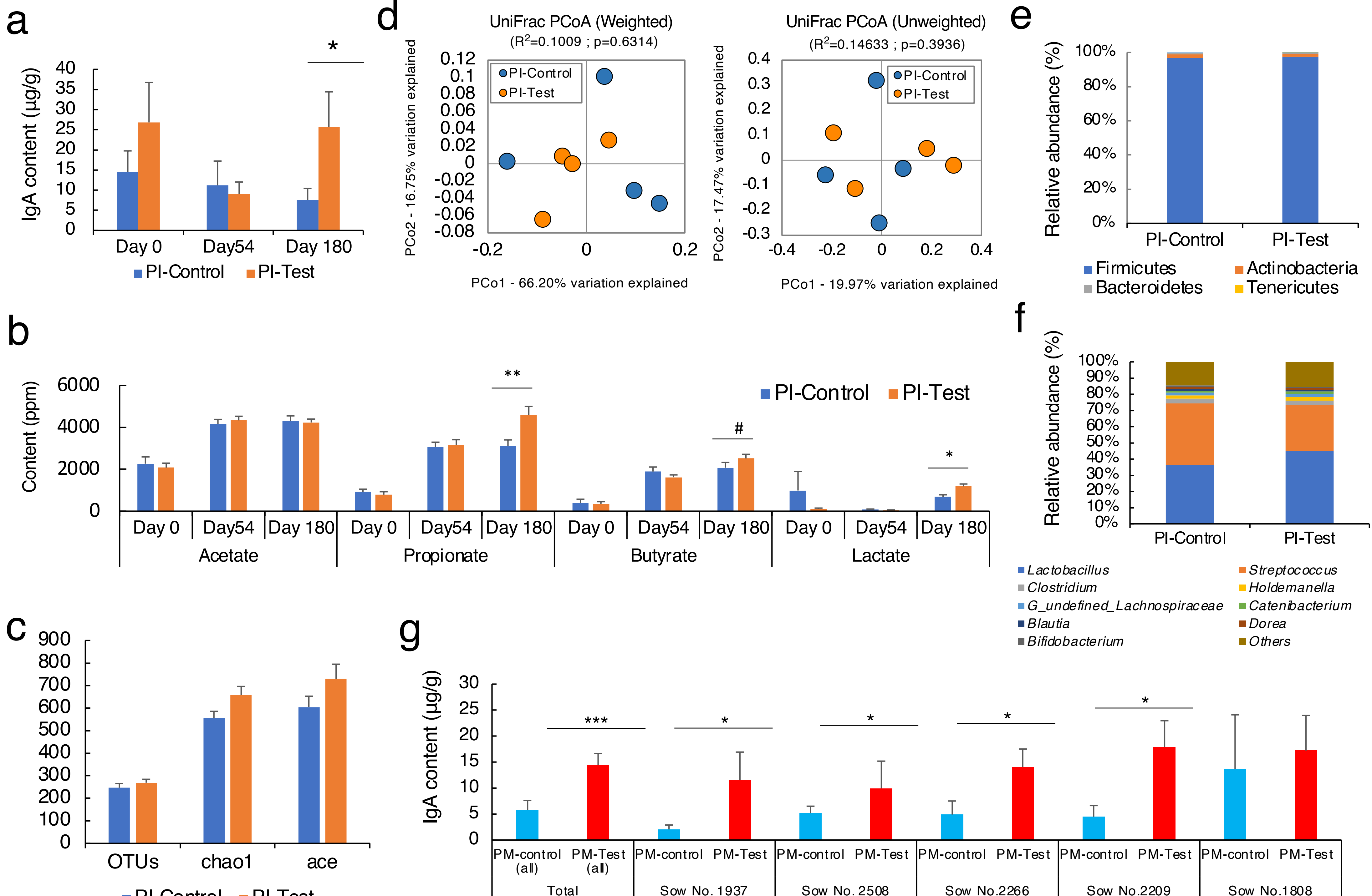

Fig.S6

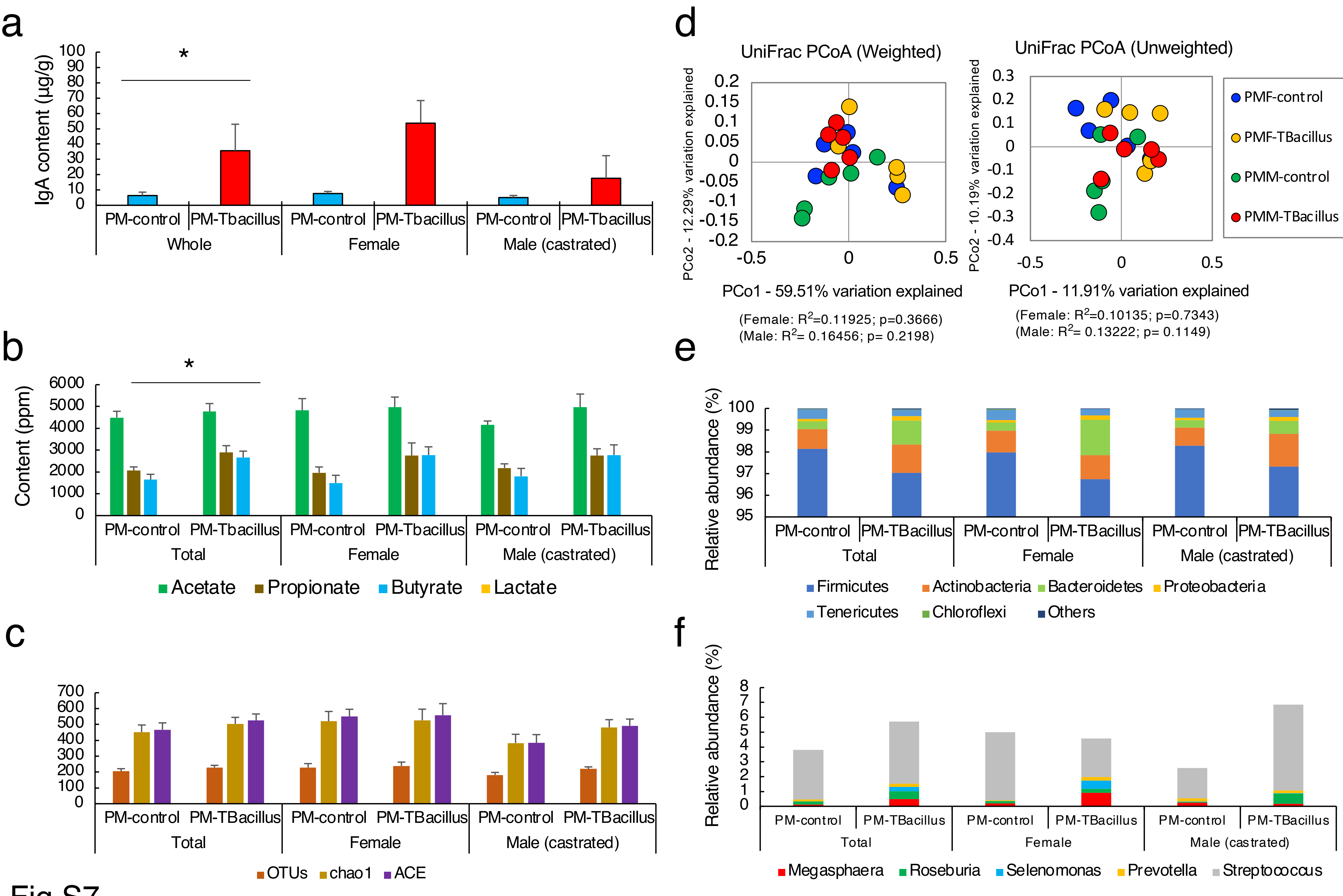

Fig.S7

a

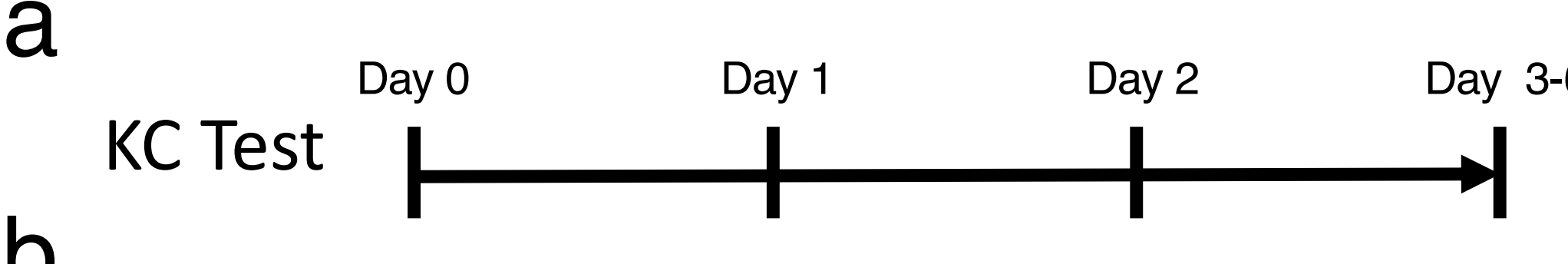

b

| Item | Non-treatment | *Caldibacillus* sp. treatment | | | | |
|---|---|---|---|---|---|---|
| | Normal condition | | Occurance of diarrchea before test | | | |
| Individual No. / Sex | 1297/♀ | 1294/♀ | 1293/♀ | 1295/♀ | 1296/♀ | 1298/♀ |
| Duration of test (Day) | 4 | 4 | 4 | 4 | 6 | 4 |
| Cow No. | [1176] | [1213] | [1141] | [1188] | [1125] | [1190] |
| Body weight of calf (Birth) | 33 | 32 | 29 | 38 | 28 | 35.5 |
| Beginning age of test (Day) | 78 | 77 | 103 | 74 | 116 | 74 |
| Beginning body weight (kg) | 83.1 | 99.5 | 107.9 | 105.8 | 117.1 | 86.1 |

c

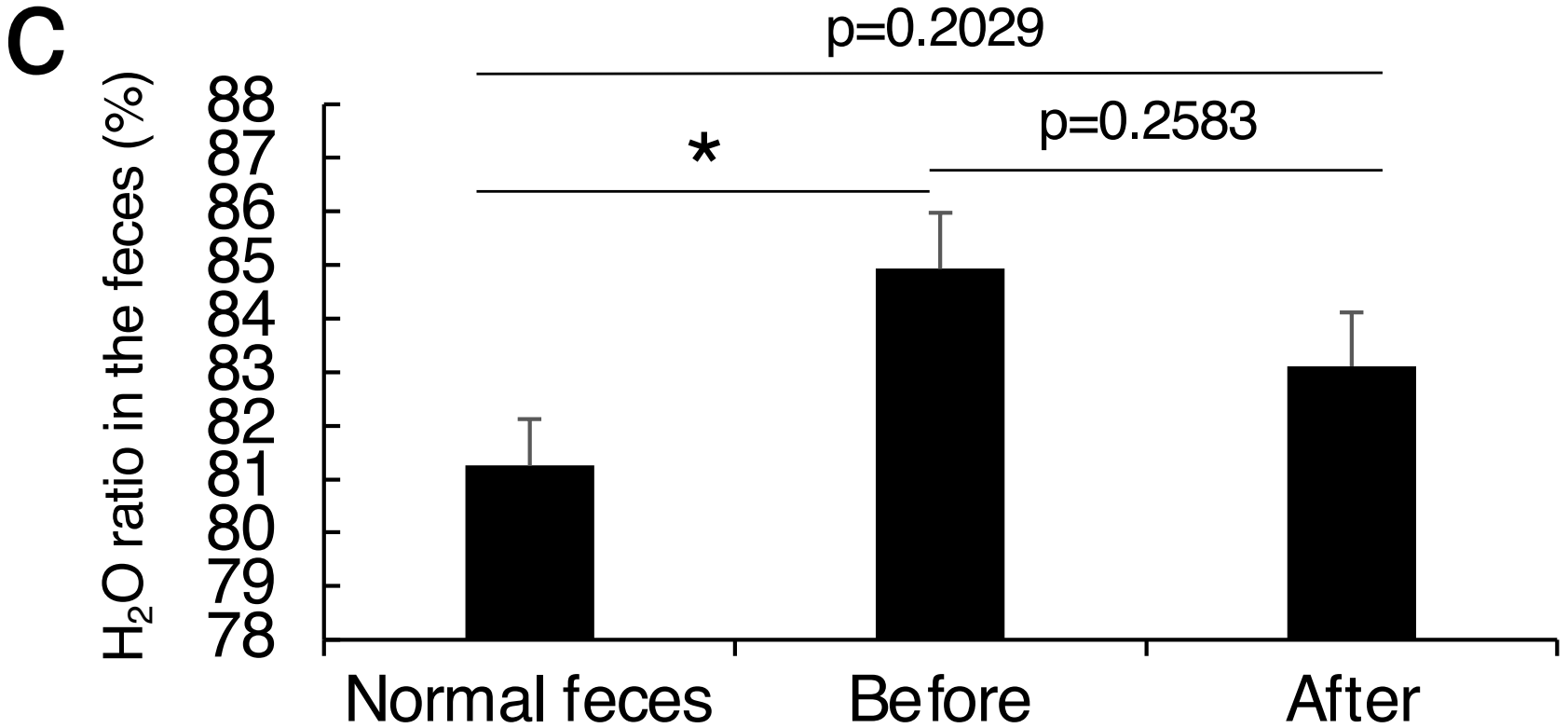

d

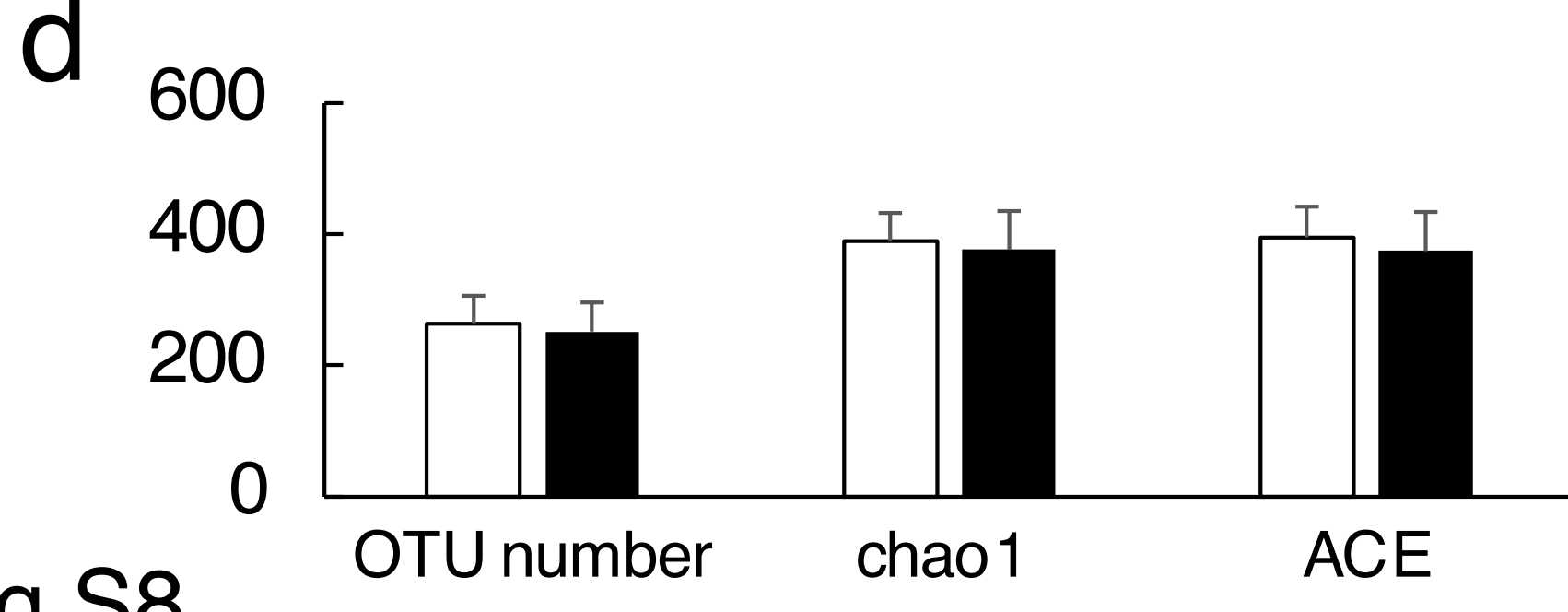

e

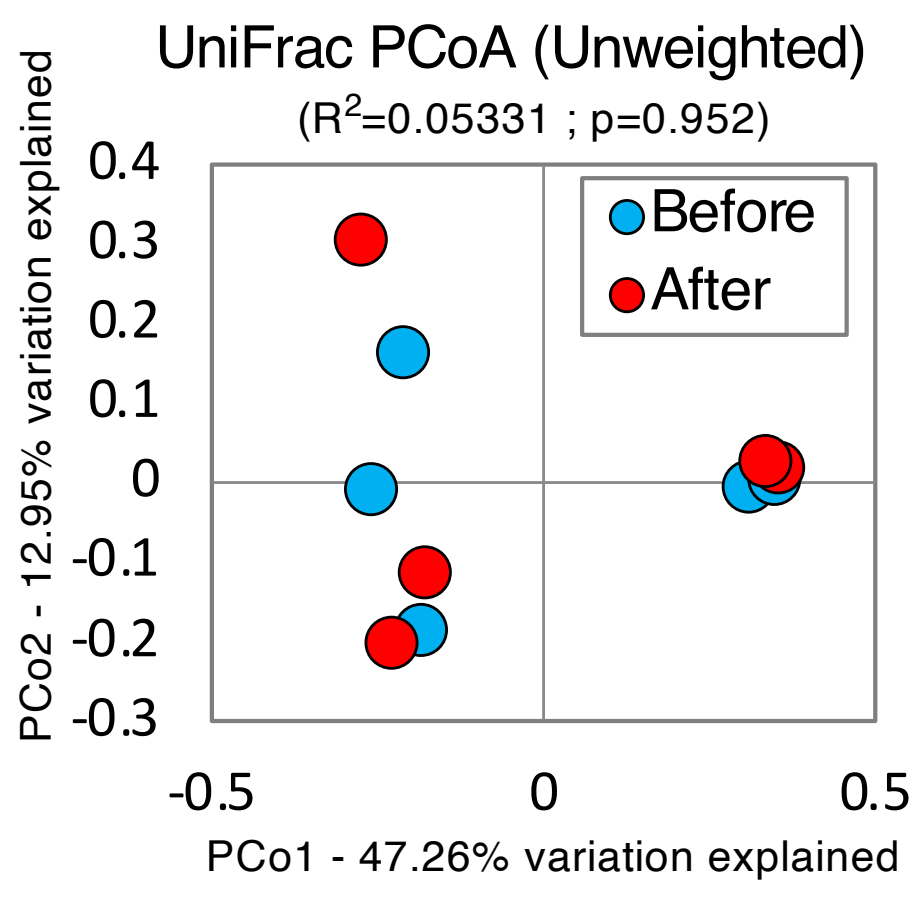

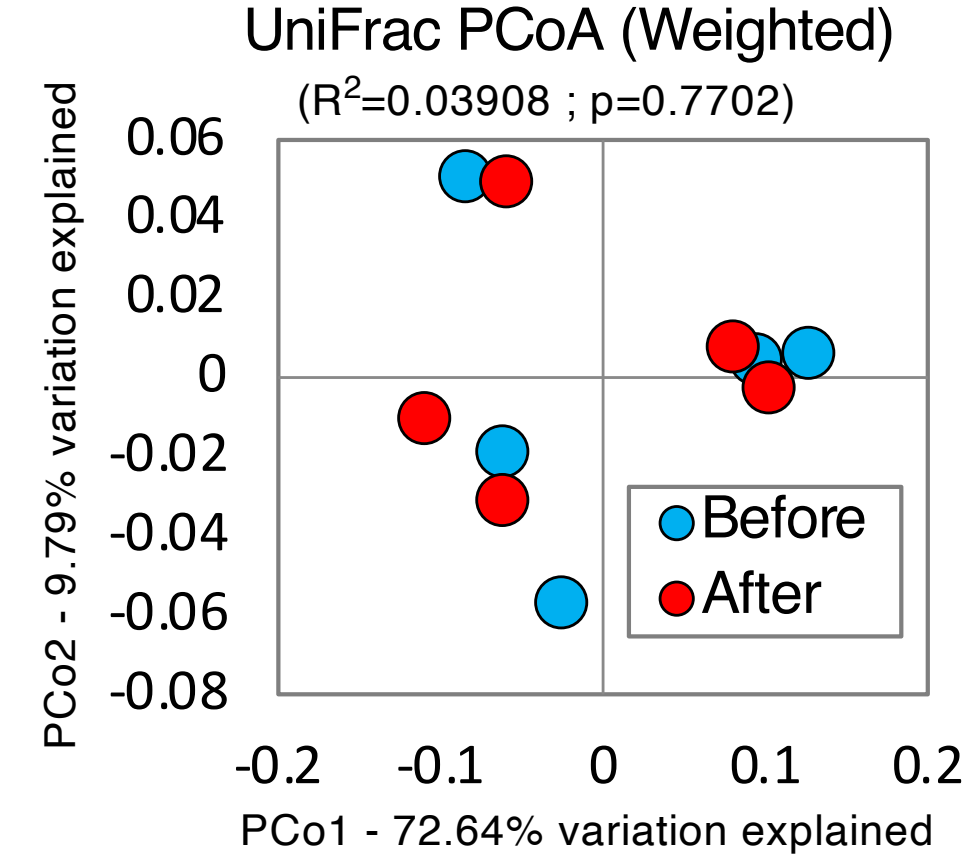

f

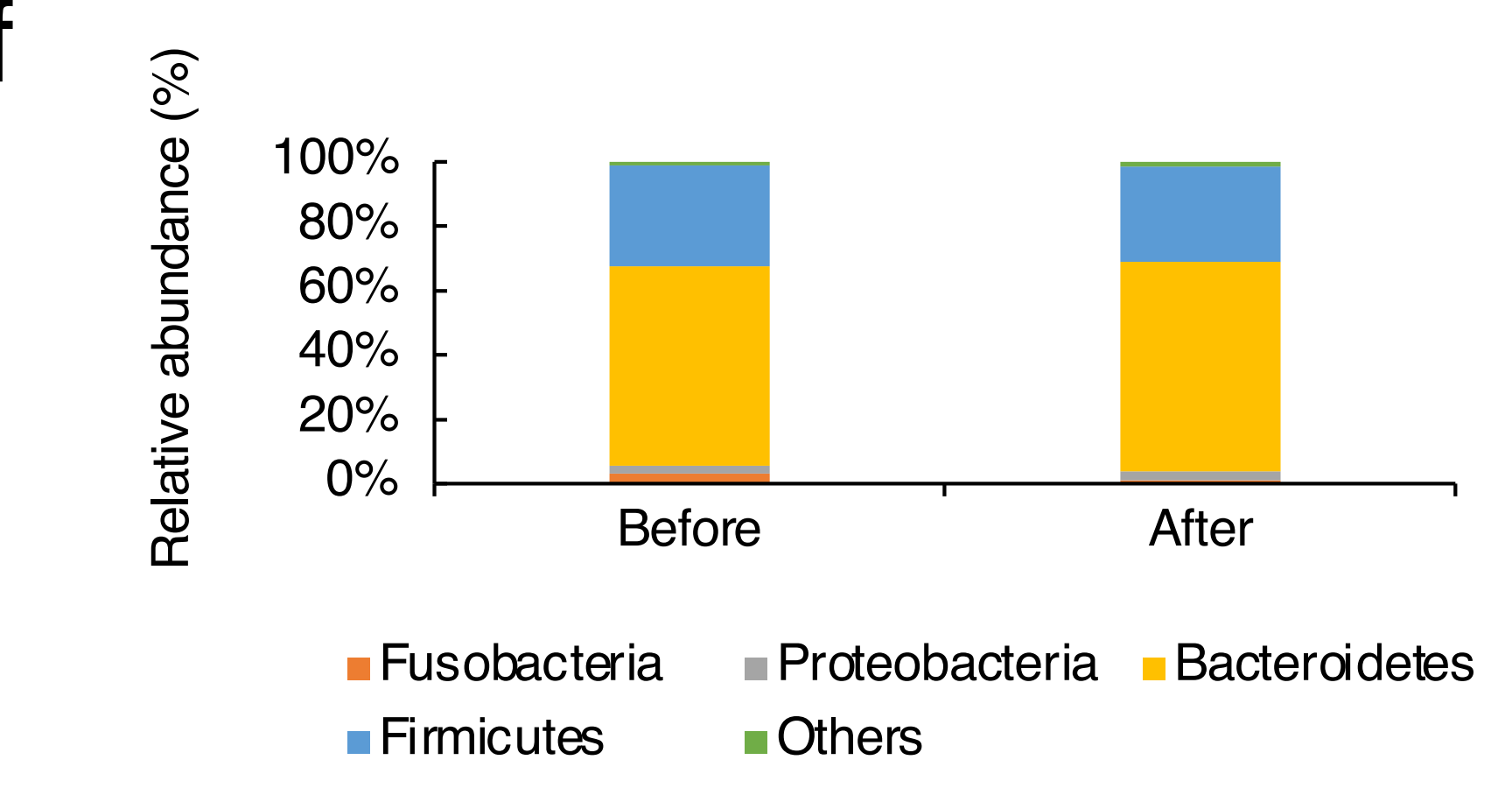

g

| Origin | Pathogenic factors | | Calf no. | | | | | |
|---|---|---|---|---|---|---|---|---|
| | Gene name | Gene sequence | 1297 | 1294 | 1293 | 1295 | 1296 | 1298 |
| Pathogenic *Escherichia coli* | saa | SAA | - | - | ○ | - | - | - |
| | | Vt com | - | - | ○ | ○ | - | - |
| | stx | VT1 | - | - | ○ | ○ | - | - |
| | | VT2 | - | - | - | - | - | - |
| | stx1 | STx-1 | - | - | ○ | ○ | - | - |
| | sta | STa | - | - | - | - | - | - |
| | f17 | F17 | - | ○ | - | - | - | - |
| | lt | LT | - | ○ | - | - | - | - |
| *Salmonella* sp. | inv A | | - | - | - | - | - | - |

Fig.S8

a

b

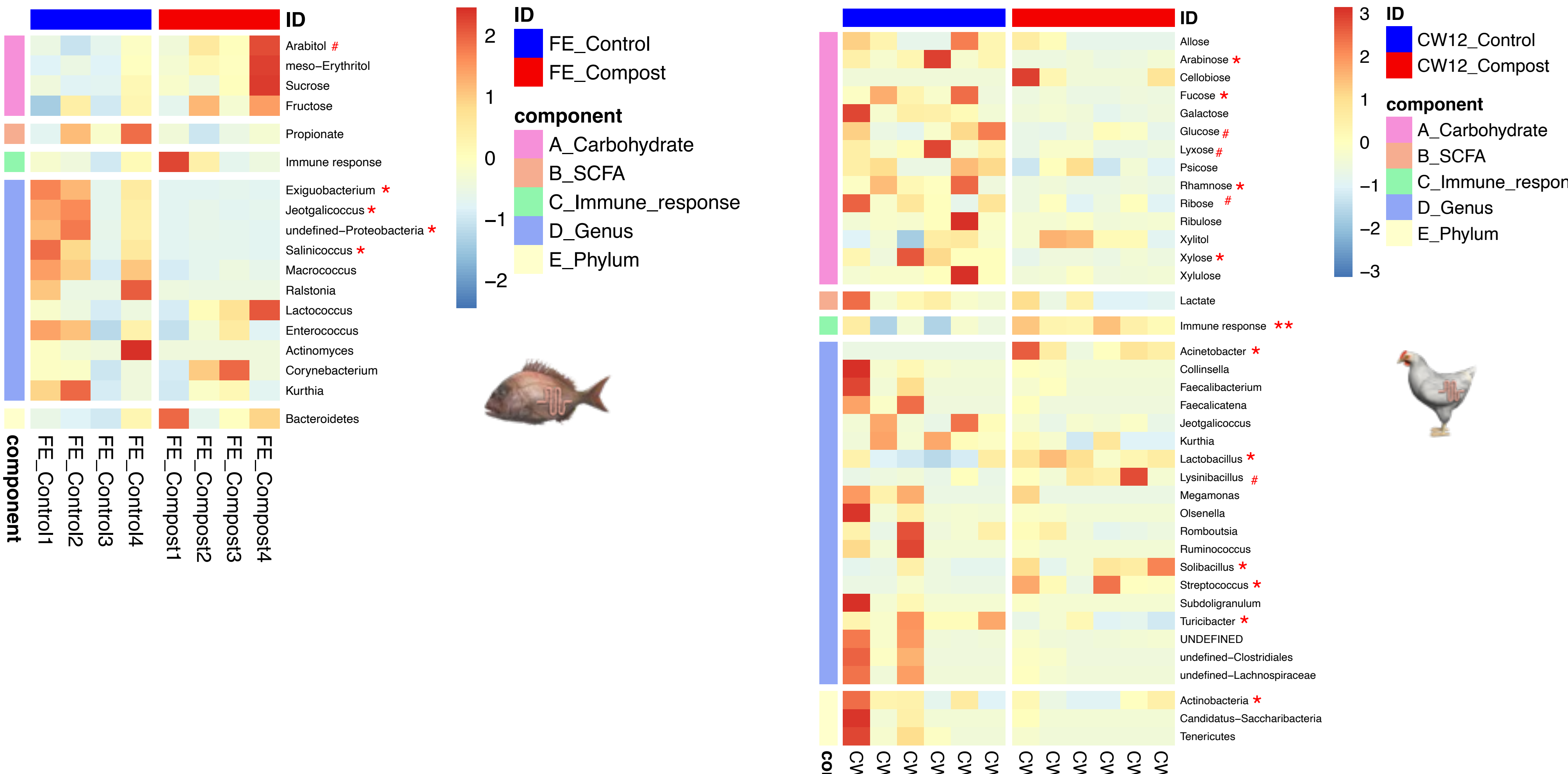

Fig.S9

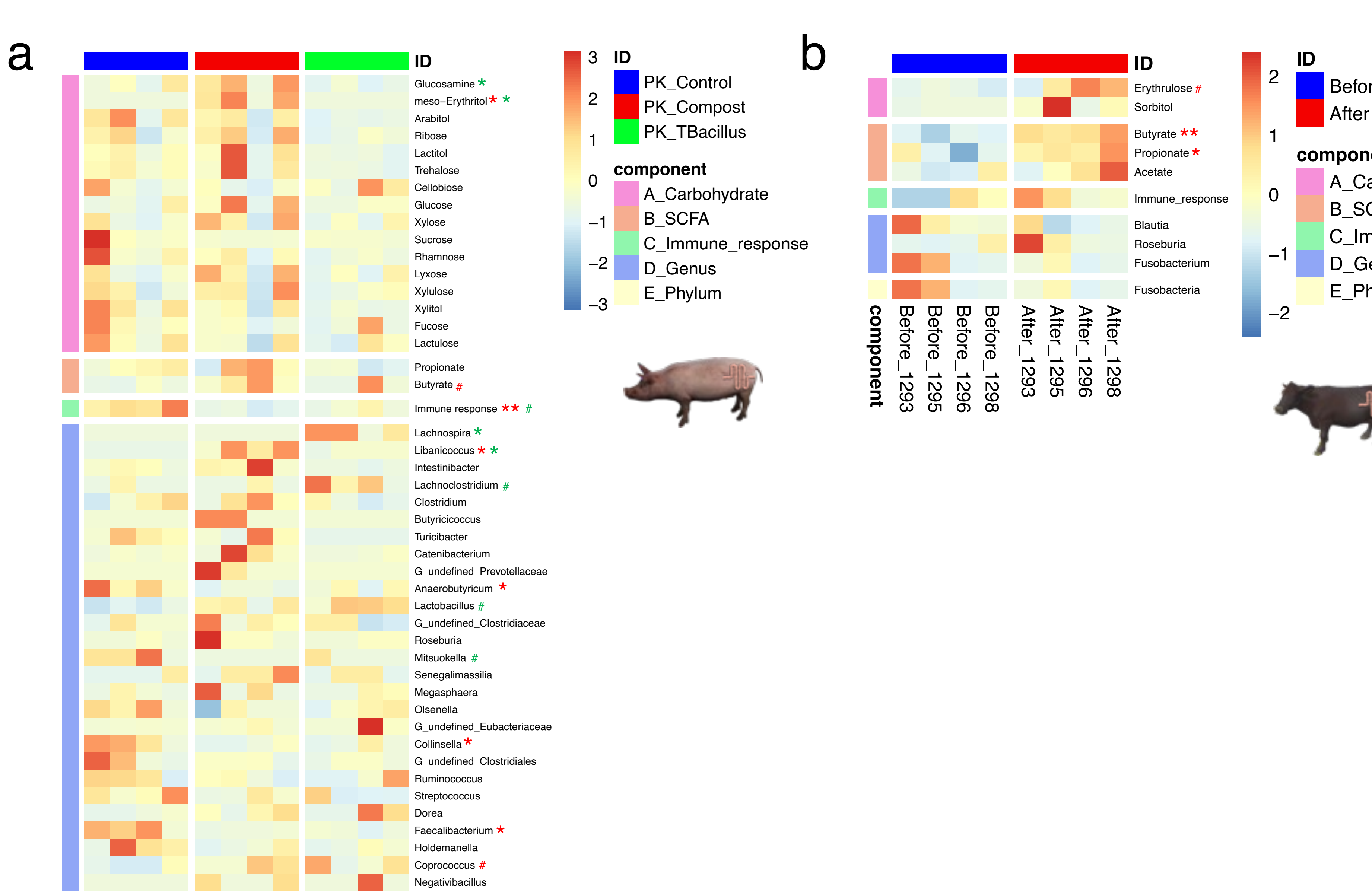

Fig.S10

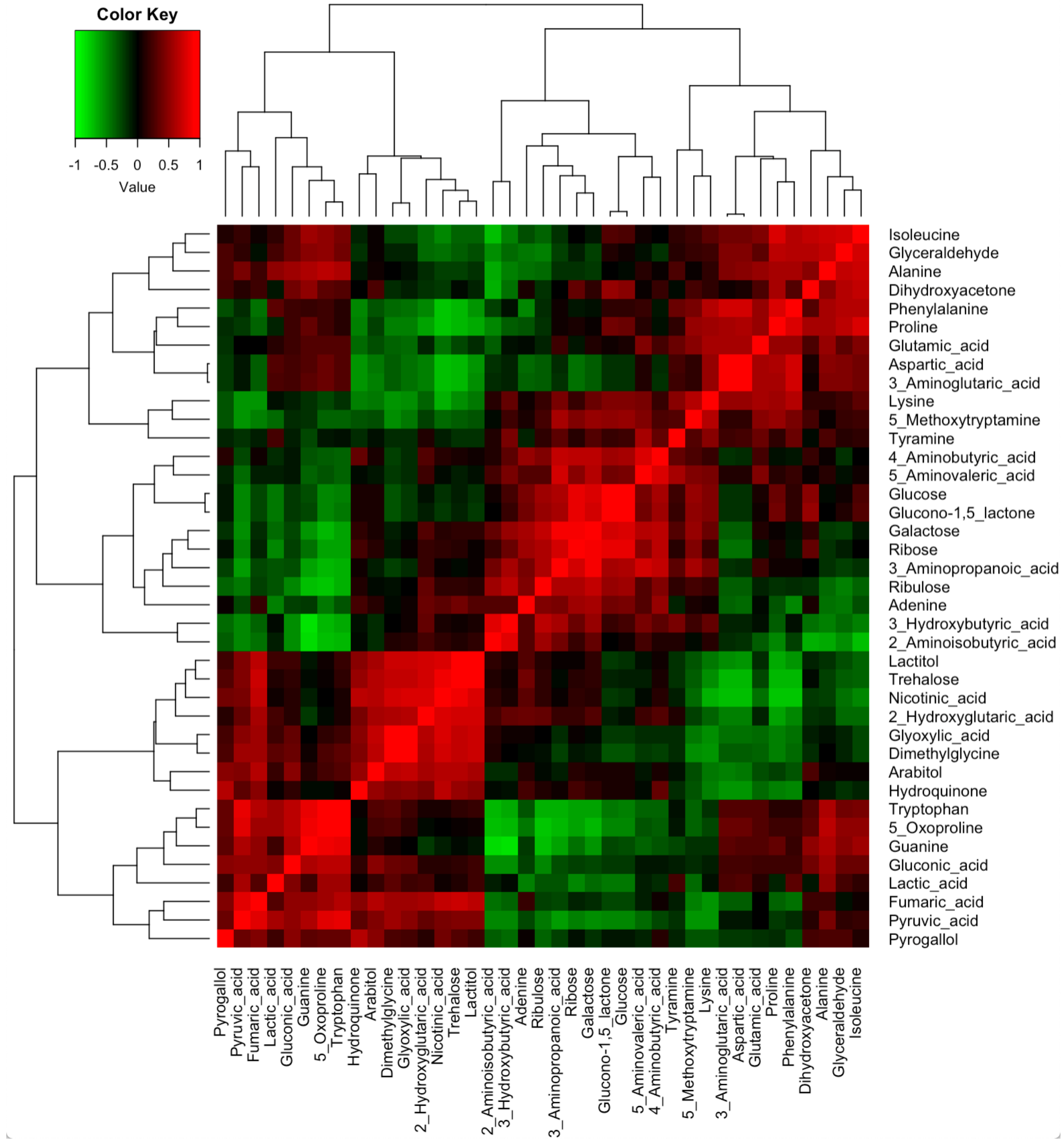

Fig. S11

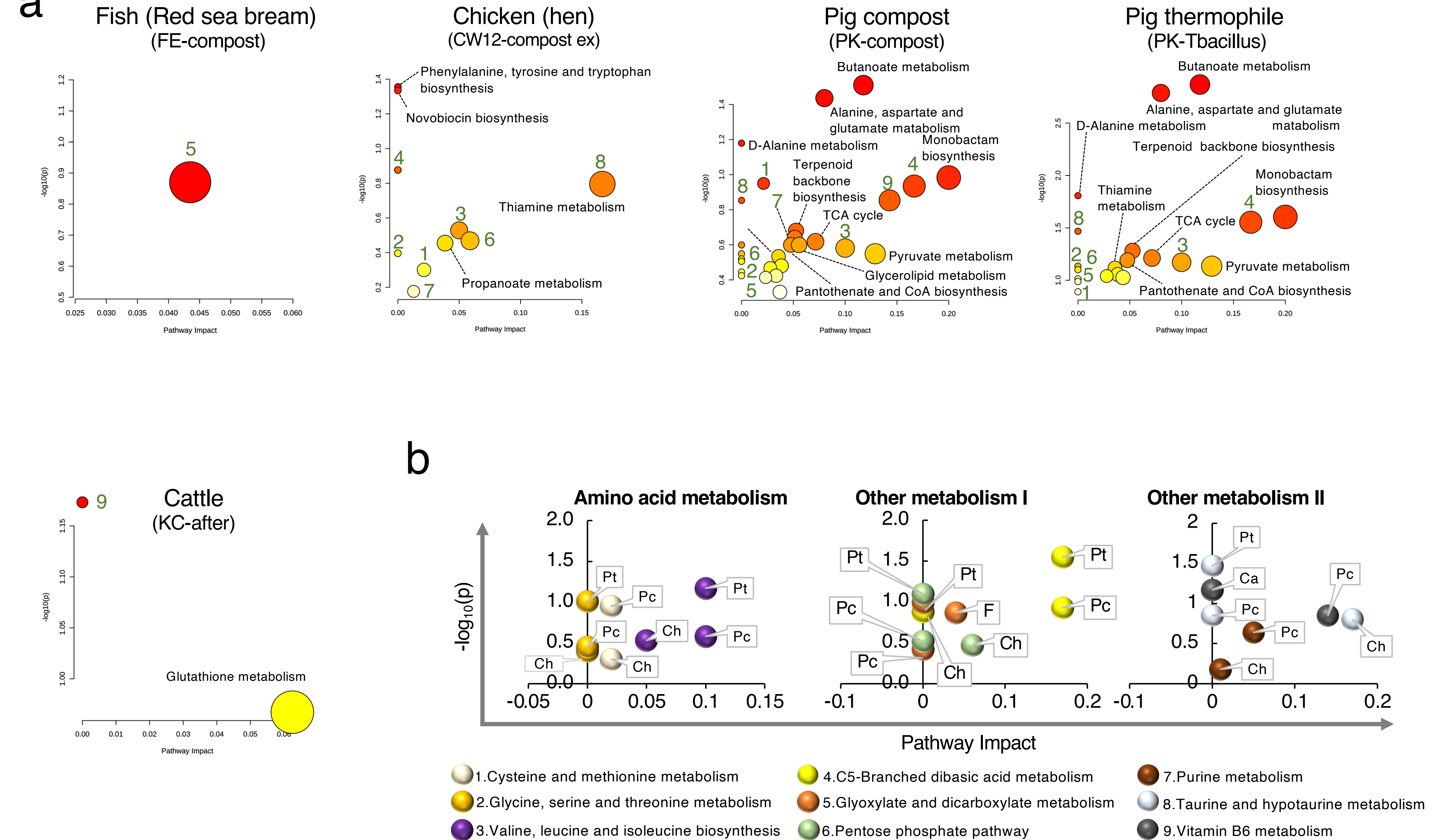

Fig. S12

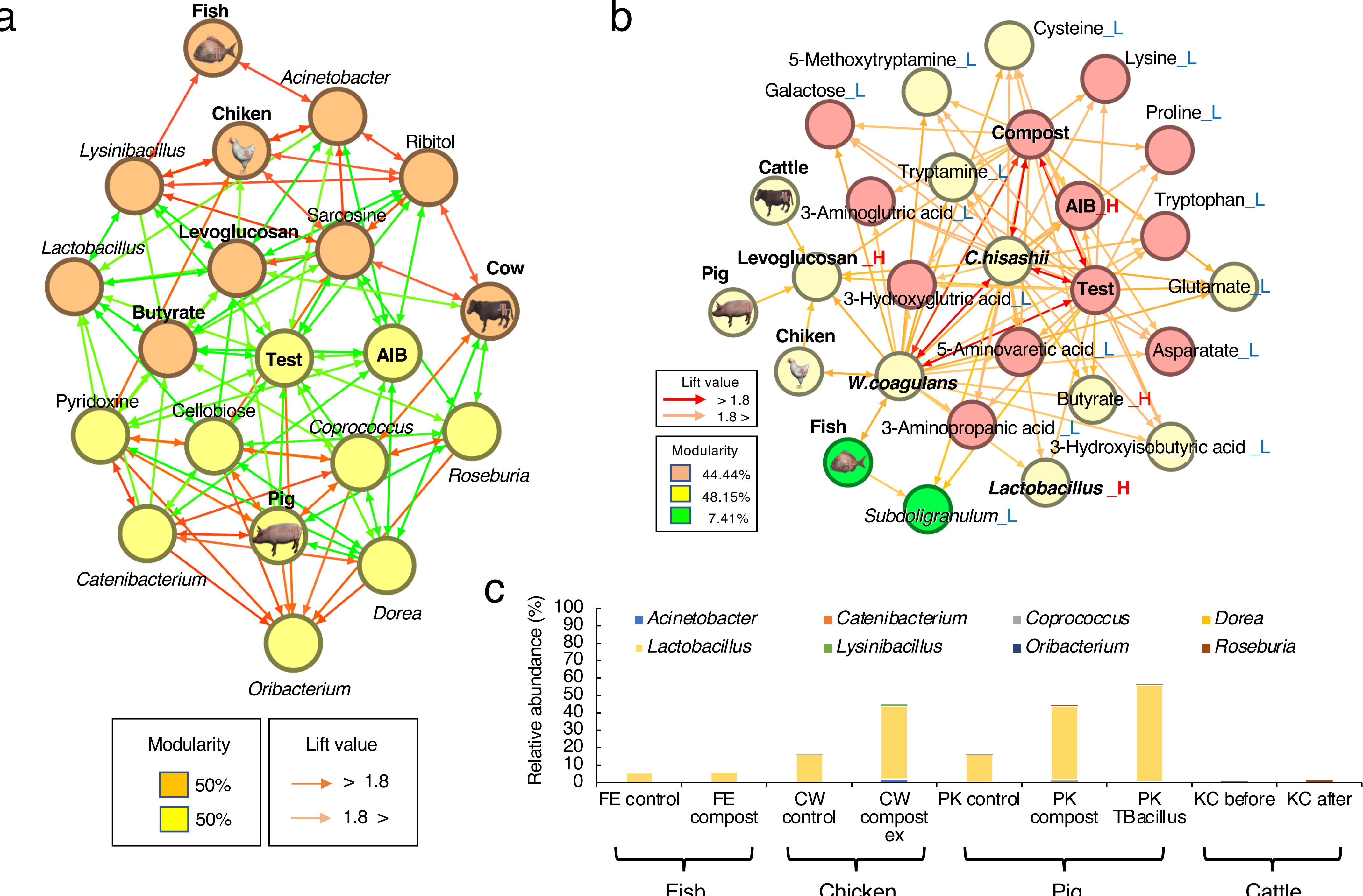

Fig.S13

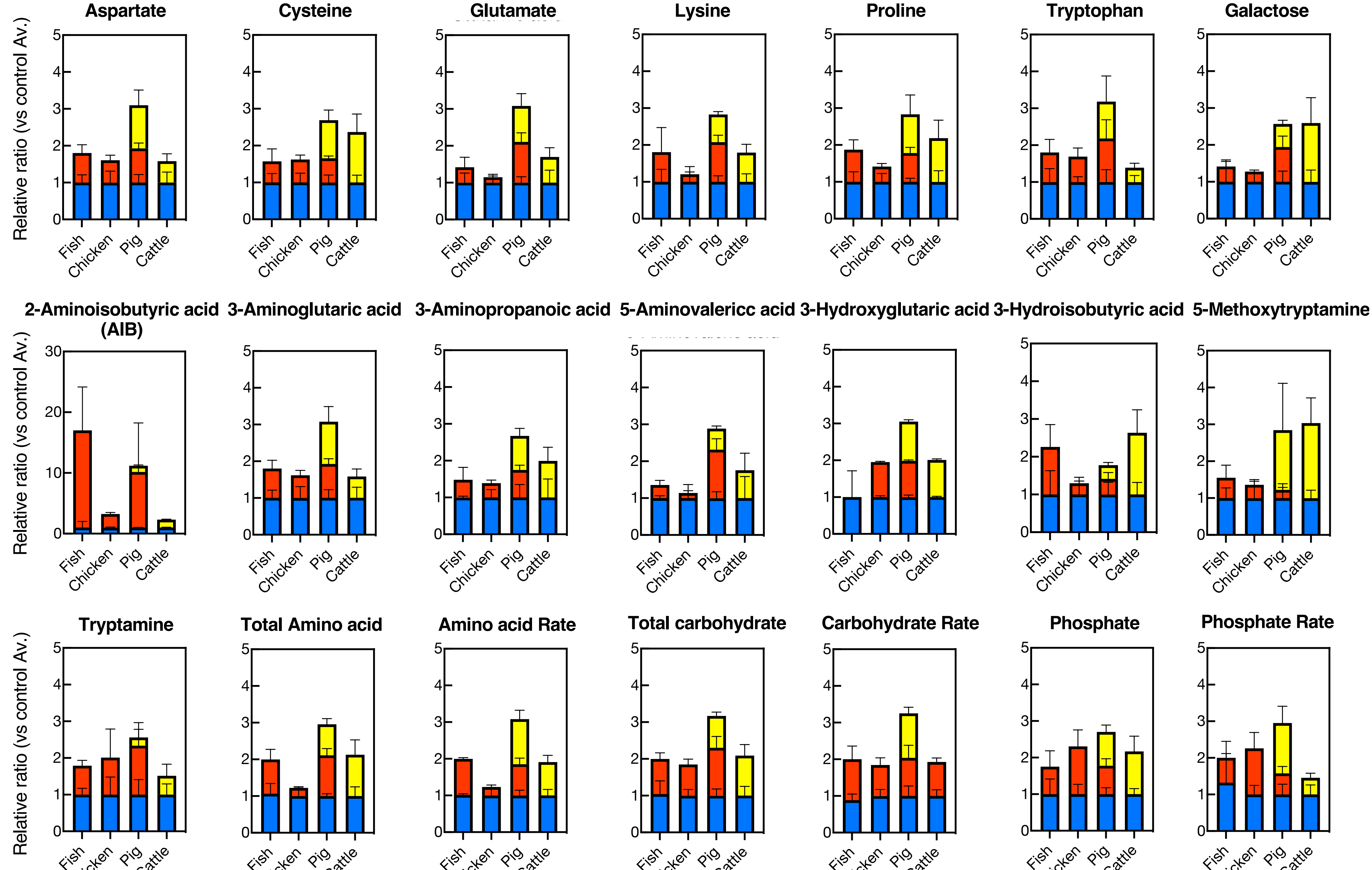

Fig.S14

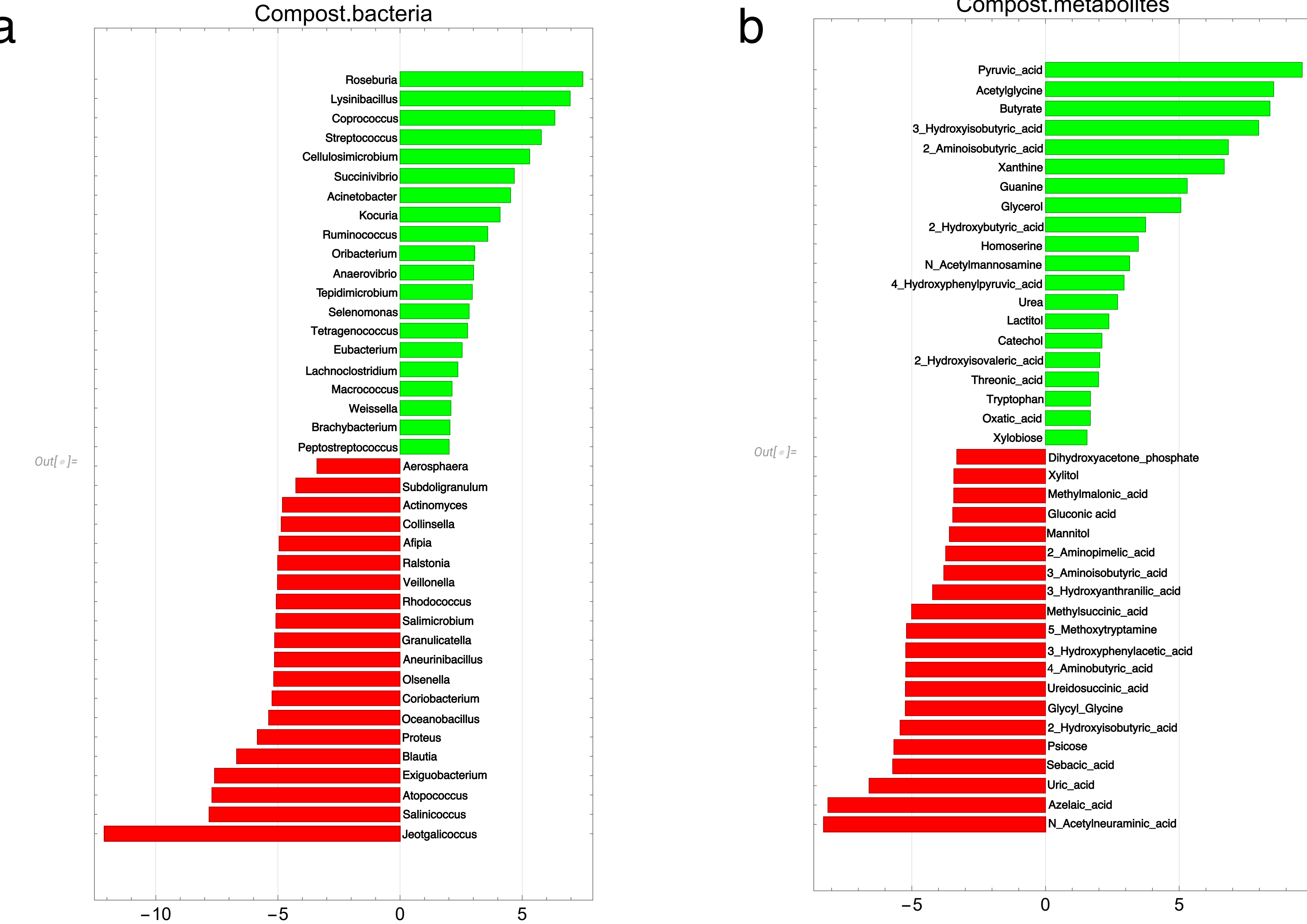

Fig.S15

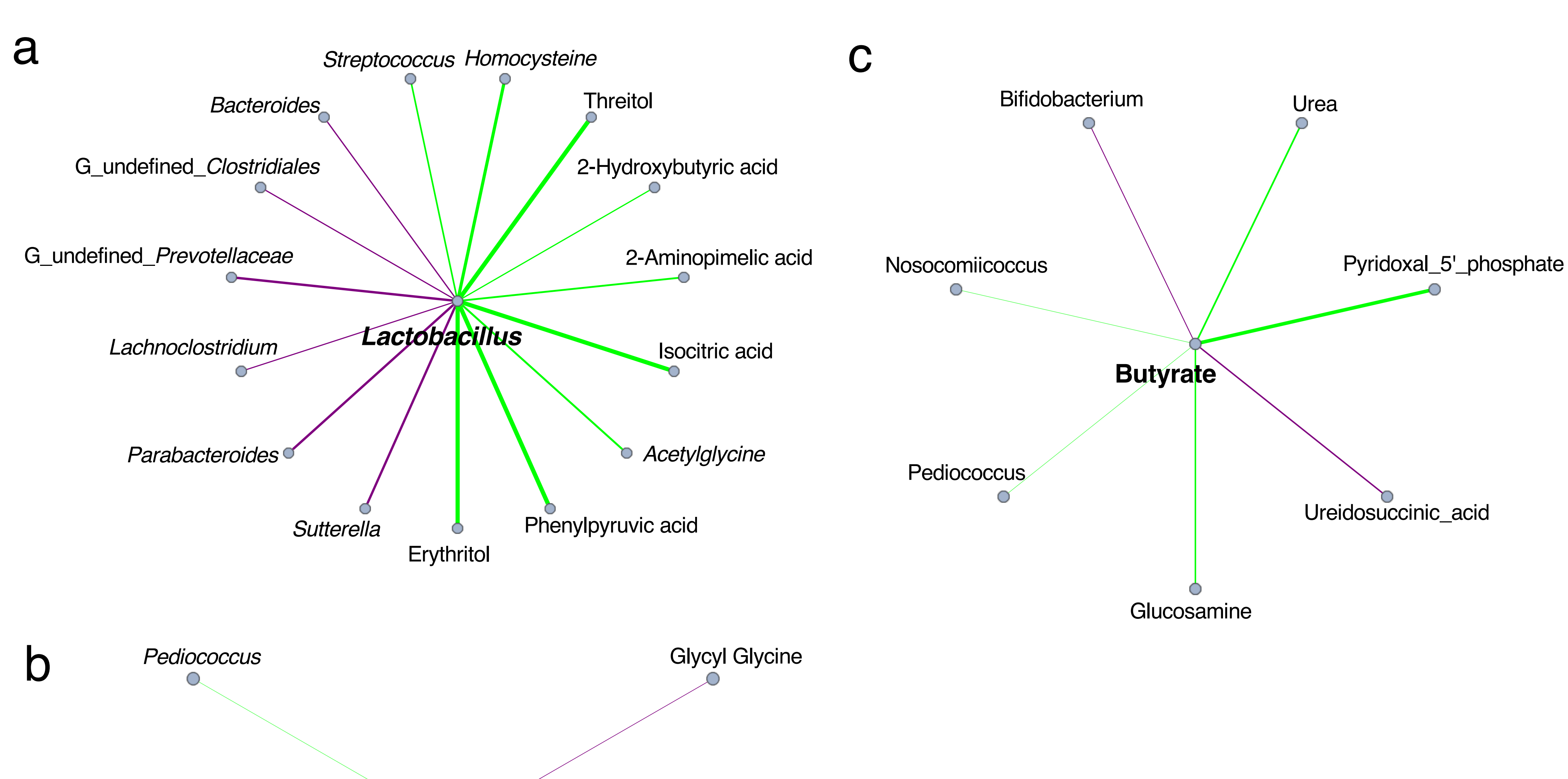

Fig.S16

a

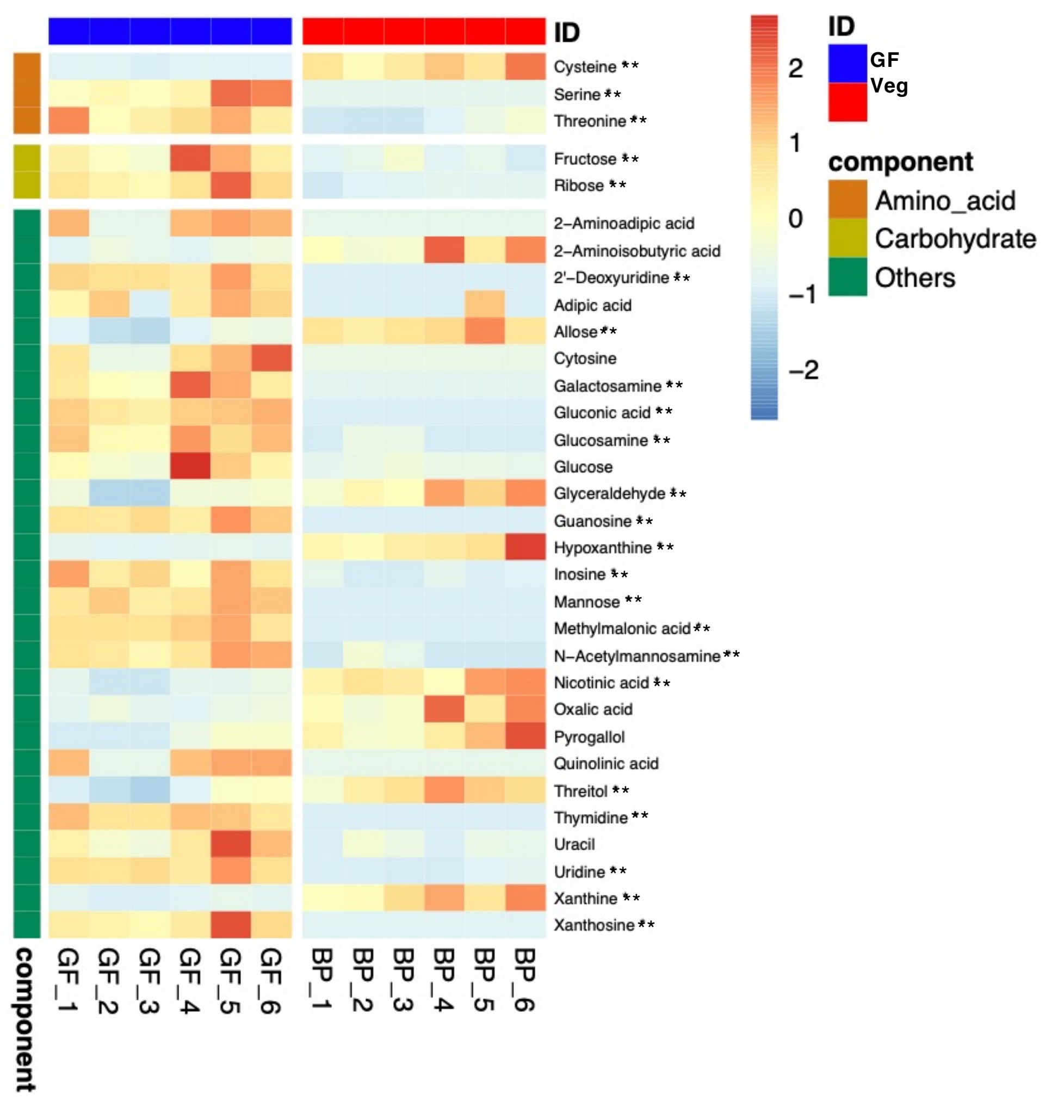

b

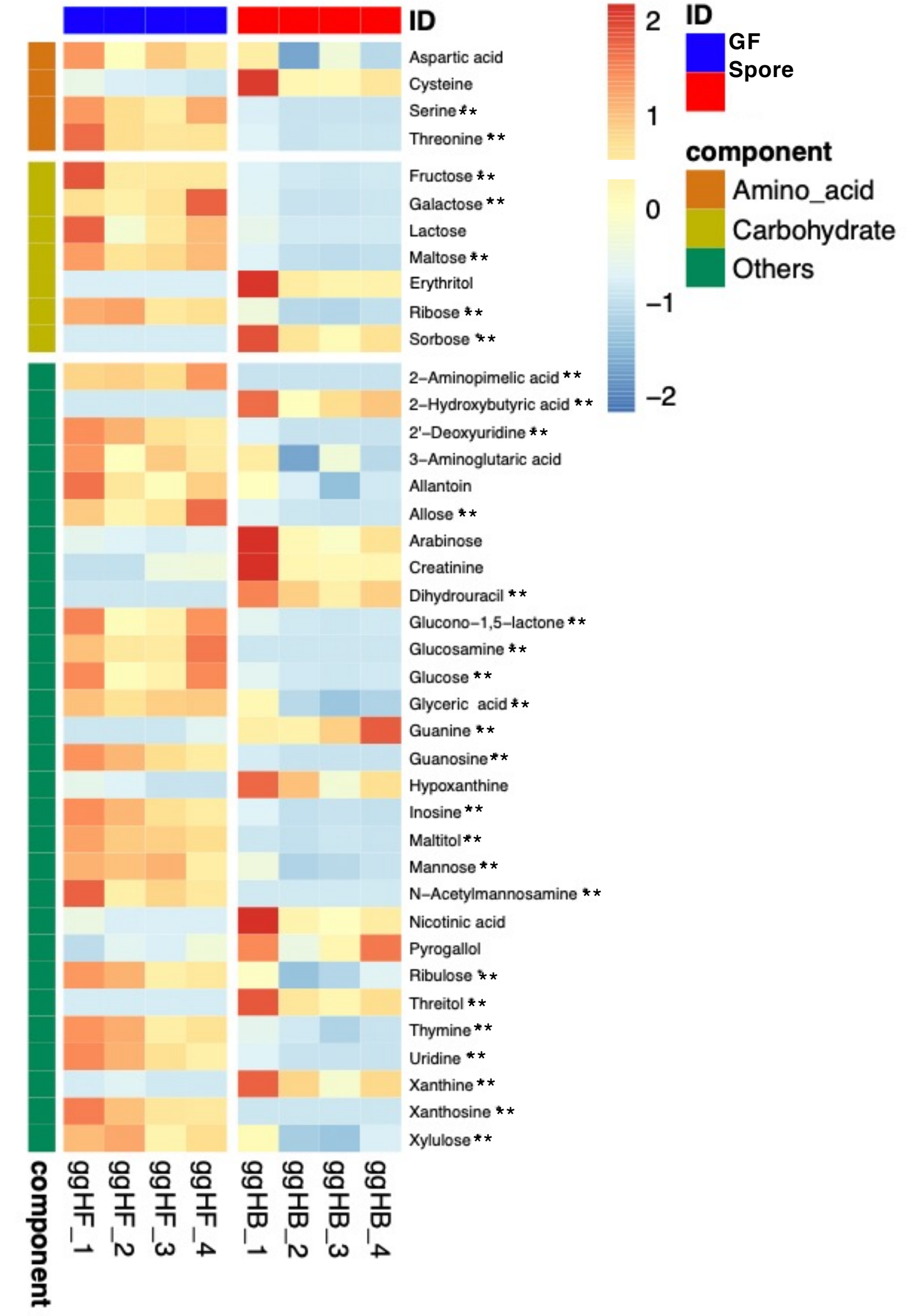

Fig.S17

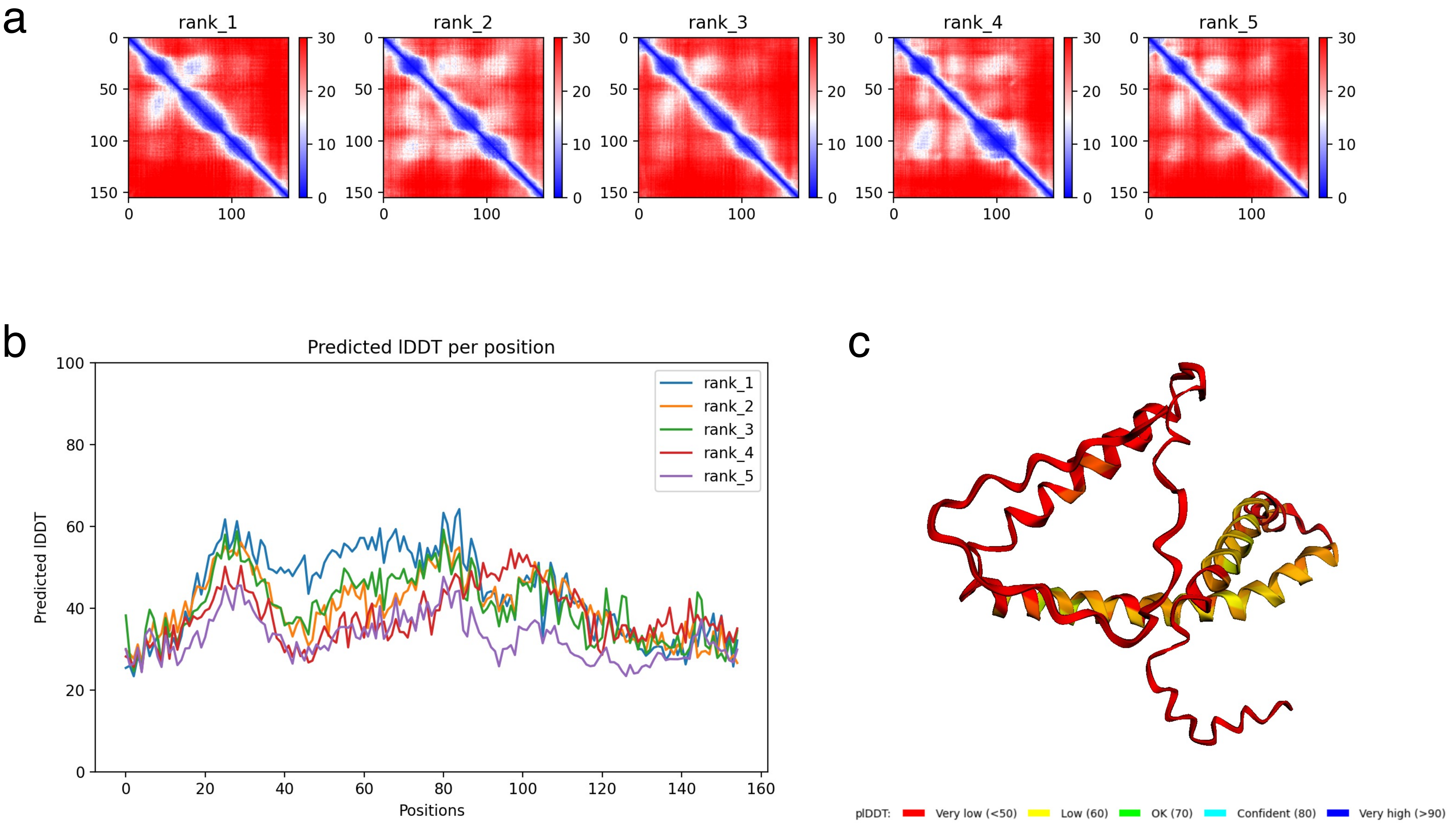

a

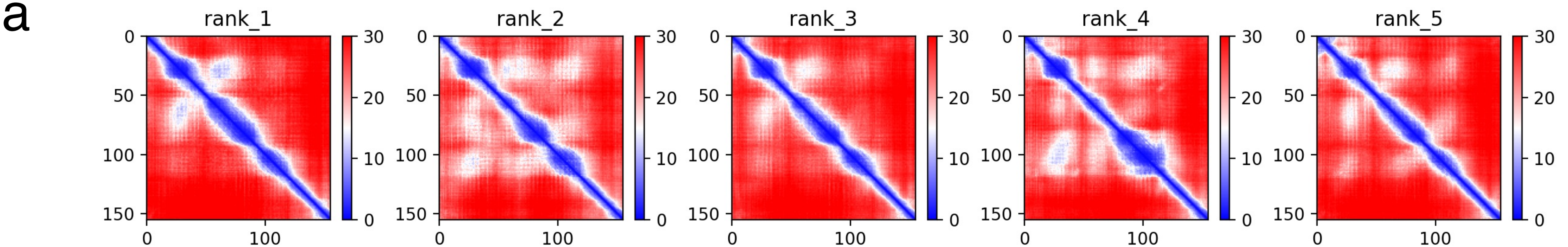

b

c

Fig.S18

# Supplementary Methods

Title: **A novel sustainable role of compost as a universal protective substitute for fish, chicken, pig, and cattle.**

### *Preparation of the compost as a feed additive and the compost-derived* **Bacillaceare**

The compost fermented of the marine animal resources (MAR compost) was prepared using an aerobic repeated fed-batch fermentation system at high temperatures (approximately 75°C) (fermentation-associated self-heating) as previously described[1] (Miroku Co. Ltd. and Keiyo Gas Energy Solution Co. Ltd., Japan). The powdered compost and/or its extract were used as appropriate for feeding conditions of experimental animals. The extract was prepared to be diluted 1/100 with potable water (as vol./vol.) and incubated under aerobic conditions at 60°C for at least 10 h (this solution was used as the compost extract)[2-7]. The two thermophiles, compost-derived Bacillaceae, were prepared by Sermas Co., ltd., Japan. Firstly, *C.hisashii,* which was first isolated from the MAR compost as the *Bacillus thermoamylovorans* N11 strain[7], registered as *B. hisashii* (Type stain N-11$^{T}$ =NRBC 110226$^{T}$ =LMG 28201$^{T}$)[8], registered as the accession number NITE BP-863 given by the international depositary authority in National Institute of Technology and Evolution (NITE) in Japan (Sep 26, 2014), and recently reclassified to *C. hisashii*[9], were cultivated as previously described[7]. Secondly, N16 strain, which was first isolated from the MAR compost as the *B. coagulans* N16 strain[7], registered as the accession number NITE BP-02066 given by the international depositary authority in NITE (July 1, 2015), and recently reclassified to *Weizmannia coagulans*[9]. In CW and CCR test (Fig.S.3de), the compost extract concentration was more reduced to at least 1/1000 so far (diluted compost extract), and the amount of the probiotic candidates *C.hisashii* N-11 stain and *W.coagulans* N-16 stain were adjusted. A compost extract comprising a 1:1 mixture of two stains and adjusted to $10^{2-3}$ CFUml$^{-1}$ in portable water was orally administered *ad libitum*. In the PK and PF swine tests (Fig.S.3gh), *C. hisashii* was adjusted to $10^{2-3}$ CFUg$^{-1}$ in total feed in the compost group. Only *C. hisashii* strain in the TB group of the swine test was adjusted to $10^{2-3}$ CFUg$^{-1}$ in total feed. As an additional excipient, defatted rice gran in the PK test and skim milk in PF test was used as vehicle at 0.1 % or 0.2% total feed, respectively. In the PI test (Fig.S.3i), *C. hisashii* in the diluted compost extract was adjusted to $10^{2-3}$ CFUg$^{-1}$ as a material in liquid feed until first Day 57 of the test ; thereafter, only the *C. hisashii* N-11 strain (0.1%) with defatted rice gran as an additional excipient was administered. In the PM test of swine test (Fig.S.3j), only the *C. hisashii* strain, adjusted to $10^{2-3}$ CFUg$^{-1}$ in total feed, with defatted rice gran as an additional excipient (0.1%) was administered. Similarly, in the KC cattle test (Fig.S.3k), only the *C. hisashii* strain ($10^{2-3}$ CFUg$^{-1}$ as total feed) with potato starch as an additional excipient was administered.

### *Fish management*

Artificial infection tests with *Edwadsiella tarda* were performed with the red sea bream *Pagrus major* as a fish model in the laboratory water tanks circulating in one way with sea water from two regions, Shizuoka (Chubushiryo, Japan) and Ehime (Ehime Univ., Japan). In the Shizuoka test (FS-Test in Fig.S.3a), the three groups of red sea bream (n=15 per group in 0.5 t capacity of a round tank) were fed with commercially available extruded pellet (EP) feed (Tai Next EP; Chubushiryo, Japan) supplemented with 0 %, 1 %, and 5 % (weight/weight) of the water extract of thermophile-fermented compost[2-6,8,10]; these treatments were named as follows: FS-control, FS-extract 1%, and FS-extract 5%. The temperature of the water was maintained at approximately 22 °C during February and May. After these continuous conditions for two month, serum samples from some of the fish (n=5 per group) was analysed for complement activity test[11], and artificial infection of the fish (n=10 per group) *i.p.* with 0.3ml (conc. 4.3 $\times 10^7$ cfu/ml) *E.tarda* (E05-92; Chubushiryo, Japan) was performed for survival analysis. In the test in Ehime (FE-Test in Fig.S.3b), the four groups of red sea bream (n=40 in 1.0 t capacity of round tank) were fed with commercially available extruded pellet (EP) feed (Madai EP super No.2; Marubeni nisshin Feed, Japan) added with 0 %, 1 %, and 5 % (weight/weight) of thermophile-fermented compost powder or 1% (weight/weight) of the water extract of thermophile-fermented compost[2-6,8,10], named in order as follows: FE-control, FE-1% compost, FE-5% compost, and FE-1% compost ex. Thermophile-fermented compost powder was pasted to 1 kg of EP feed using 3 g of a non-ionic emulsifier, Apifac (Kohkin Chemical, Japan). The temperature of water was maintained around at 16.5 °C during Jan and Feb. After these continuous conditions during one month, artificial infection with 0.1ml (two round tank with at least conc.1.57 $\times 10^6$ cfu / ml per each group) of *E. tarda*, a strain given from Fish disease diagnosis labo of Ainan (Nishiumi Branch, Ainan Town Office, Ainan, Ehime, Japan), to residues of the fish (a tank per blank group from 0% of power group, n=11; and two tanks per other individual group, n=12-15) was performed for survival analysis. After analysis, the fecal samples in their intestines were collected for omics analysis.

***Chicken management***

Three hen strains (Julia, Julia lite, or Boris Brown) were conventionally maintained at one experimental hen house and eleven hen farms, respectively, and utilized to analyse the physiological functions of oral administration of the compost. The three farms were as follows: CC farm (Chiba Prefectural Livestock Research Center, Chiba), CW farm (Wada farm and Nihon layer, Gifu), and CCR farms (Crest, Japan). All the chickens were orally administered commercial feeds and tap water *ad libitum* according to individual farm-specific guidelines. Artificial infection tests with *Salmonella enteritidis* were performed with the chick as a chicken model (Chiba Prefectural Livestock Research Center, Japan). In the chick test (CC Test in Fig.S.3c), the two groups of eggs (n=50 per group) of hen strain Julia were prepared and maintained at the CC farm with in-house blended feed available in the center. The chicks in the compost extract group were continuously administered approximately 5 ml of the compost extract per 50 hens per day in tap water *ad libitum*. After 13 weeks of the continuous administration, some chicks (n=7 per group) were utilized for the infection test and orally administered 0.5 ml ($3.2 \times 10^8$ cfu per chick) of a rifampicin (rif) resistant-strain of *Salmonella enteritidis* (SEZK-2a; Zen-noh Institute of Animal Health, Japan)[12]. The detection of the infection ratio was performed as following described procedure[12].

For the adult chicken tests, CW Test in the same hen house (CW farm) (Wada farm / Nihon layer, Gifu) (Fig.S.3d) and CCR Test in the different hen houses (CCR farms) (Crest, Japan) (Fig.S.3e) were performed. In the CW Test, the two groups of young Boris Brown hens (day 114 and 118 after birth as an introduction date) (CW-control group: n=11,667; CW-compost ex group:11,668 per group) were divided into two compartments of the same hen house with 20 days off from the introduction date (17 days off after birth) and maintained with in-house blended feed available on the farm. In the CCR Test, ten groups of hen houses (from day 120 after birth as an introduction date) (n=27,796-64,783 /hen house at the initial date) of the Julia lite strain were maintained with in-house blended feed available on the CCR farm. The compost extract was adjusted to contain $10^{2\text{-}3}$ cfu/ml of *C.hisashii* N-11 strain and *W.coagulans* N-16 strain as previously described as final drinking concentration[8].

***Swine management***

Two swine strains, LWD, which is crossbred with Landrace × Large White × Duroc, and Berkshire, were conventionally maintained on five types of farms and utilized to analyse the physiological functions of oral administration of the compost and/or *C.hisashii* : LWD, PK farm (Chubushiryo, Aichi), newly established pens, PF pens (Shoku-Kan-Ken, Gumma), and PI farm (Itou swine farm, Chiba); Berkshire, PM farm (Minami-nihon-chikusan, Kagoshima). All the pigs were orally administered commercial feeds and tap water *ad libitum* according to individual farm-specific guidelines. The PK Test in Fig.S.3g was also conducted on the PK farm (Chubushiryo, Aichi). On the farm, the weaned piglets were maintained with two types of feeds (piglet special commercial feeds, Chubushiryo Co., Ltd.) on days 25-71 after birth. The effects of administration of the compost and/or the *C.hisashii* N-11 strain on the weaned piglets were investigated there. In the other region (PF-Test), the effects of oral administration of only the *C.hisashii* N11 strain during days 40-61 after birth on weaned piglets, which were maintained with in-house blended feed available on the farm (Shoku-Kan-Ken, Gumma), were investigated in newly established pens (PF Test in in Fig.S.3h). In the PI Test, the effects of administration of the compost extract with the *C.hisashii* N11 strain (C-compost extract) on the weaned piglets were investigated during days 23-80 after birth and during days 80-160 (approximately day 160) after birth on the PI farm. In the administered group, growing pigs tested during days 80-160 (approximately day 160) after birth partly contained pigs not administered the C-compost extract during days 23-80 (approximately day 80). In the PM Test, the effects of administration of only the *C.hisashii* N11 strain (without the compost extract) during weaning period, days 40-130 (approximately day 130) after birth, on the growing pigs and thereafter were investigated, and the death rate and the cost of drugs during days 130-220 after birth were calculated.

### *Cattle management*

Six Japanese black calves (No.1293,1294,1295,1296,1297,1298), 88.8 ± 8.7 days of age, 103 ± 5.1 kg body weight (BW), which were born in Kuju Agricultural Research Center of Kyushu University (Oita, Japan), were used for this study. All the calves were fed milk replacer of a quality stipulated in the Japanese feeding standard for beef cattle 2008 (Agriculture, Forestry and Fishery Research Council Secretariat, MAFF 2008). Calf starter and hay were fed to all the calves *ad libitum* according to the Japanese feeding standard for beef cattle 2000 (Agriculture, Forestry and Fishery Research Council Secretariat, MAFF 2000). In brief, warm water containing 5 g of potato starch and *C.hisashii* to be adjusted to $10^{2-3}$ $CFUg^{-1}$ as total feed was administered every morning to four calves (No.1293, 1295, 1296, 1298; diarrhoea-group), which developed diarrhea, and to a calf (No.1294; normal-group), which did not develop diarrhoea. Administration of *C.hisashii* to the diarrhoea-group was performed until calves recovered from diarrhoea (from at least 4 days to 6 days), and that to the normal group was performed for 4 days. One calf (No. 1297) was offered warm water for 4 days without *C.hisashii* administration. This duration was set according to our observational data: calves were recovered from diarrhoea within 4 days following *C.hisashii* administration. Rectal feces were collected using long gloves immediately from the administered and non-administered calves on the first day and last day of the supply period.

### *Detection of infected bacteria and parasite*

In the FS-Test (Fig.S.3a), the presence of *Edwadsiella tarda* in the fish after the infection test was determined by the proliferation of the colonies on generally known agar media. In the FE-Test, the presence was checked by PCR with the primer sets as previously described[13]. To investigated the infection test of the *Salmonella* for hens in the CC-Test, the fresh caecal feces excreted on the early morning of the 4th and 7th days after infection were collected individually from all the birds and diluted to a 1/10 volume with sterile physiological saline. Then, 0.1 ml of these diluted solutions was cultivated on DHL agar containing 50 µg/ml rifampicin (rif DHL) for 24 hr at 37 °C, and thereafter, the number of resistant viable colonies and rif-resistant *Salmonella* colonies were calculated as previously described[12]. The liver and ovary tract were diluted to 1/10 volume with sterile triptic-soy-broth (TSB) (Nissui, Japan), and thereafter the number of the rif-resistant *Salmonella* colonies in organ solution were calculated by cultivation on rif DHL as the procedure for cecal feces. In order to evaluate the presence of coccidium, the numbers of oocysts per gram of feces (OPG values) were detected as previously described[14].

In order to estimate the cause occurring diarrhea of cattle, pathogenic genes of *Escherichia coli* and *Salmonera enteriridis* were detected. Stx1, The primer sets of Sta[15], LT[15], F17[15], SAA[15], Vtcom[16], VT1[16], and VT2[16], as targeted genes for the siga toxin of *E.coli*, were used as previously described[15,16]. As targetted gene for S.enteridis Salmonera (*invA* gene) One Shot PCR kit (Takara Co., Ltd., Japan)

***Preparation of fecal samples***

All the fecal samples obtained for the following omics analyses were transported at -20 °C and thereafter stored at −60 °C - −80 °C until analyses. The frozen feces were utilized for the following omics analyses; the sequencing of bacterial 16S RNA gene and metabolome analyses as the following procedure. As the metabolome analyses, HPLC Prominence (Shimazu, Japan) and chromatography-tandem mass spectrometry (GC/MS/MS), GCMS-TQ8030 triple quadrupole mass spectrometer (Shimazu, Japan), were basically performed. In addition, the excreted feces except cecal feces in the CC-Test, PS-Test, PK-Test, PF-Test, and PM-Test were analyzed their organic acids as SCFA and lactate by a capillary electrophoresis method as previously described[4,5] (9,10).

***Detection of thermostable bacteria from feces***

In order to detect the bacteria from the compost in FE Test, the colonies were cultivated on nutrient agar (Nissui co., ltd.). The colonies were directly amplified by a primer set (341f-907r), and the resultant PCR fragments were sequenced as previously described[1].

***Detection of immune response and the related defensive response***

In order to speculate the defensive response, the extent of recovery against anesthesia and immune response were determined. 2-methylquinoline, quinaldine as an anesthesia, was prepared in Chubushiryo based on the manufactured protocol. The recovery time of fish administrated with it were detected to speculate recovery function of the liver against anesthesia. The sera of the red breams were obtained, in order to speculate the immune activities, and the complement activities of the sera were determined by the previously described methods[17]. In the other animals, fecal immunoglobulin A (IgA), a mucosal immune activity as non-invasive method, was determined using an enzyme-linked immunosorbent assay (ELISA) kit according to the manufacturer's instructions (Bethyl Laboratories Inc., Montgomery, TX, USA)[8].

***DNA preparation from fecal samples***

Total DNA was prepared from the feces according to the previous report[18] with some modification. Approximately 100 mg of lyophilized feces were disrupted with zirconia beads using Micro Smash MS-100 (Tomy Seiko, Tokyo, Japan), and suspended into 600 μL of 10 mM Tris-HCl, pH 8.0, and 1 mM EDTA (TE). We transferred 475 μL fecal suspension into a 1.5-mL microcentrifuge tube. This suspension was incubated with 15 mg/mL lysozyme for 1 h at 37°C. Achromopeptidase (FUJIFILM Wako Pure Chemical Corp., Osaka, Japan) was added at a final concentration of 2,000 units/mL, and further incubated for 30 min at 37°C. Furthermore, and finally 25.7 μL of proteinase K and sodium dodecyl sulfate were added at a final concentration of 1mg/mL and 1%, respectively, and then the mixed solution was incubated at 55°C for 1 h. These enzymatic treatments were carried out with shaking (1,000 rpm). The resulting lysate was treated with phenol/chloroform/isamyl alcohol, and DNA was precipitated with ethanol. DNA pellet was rinsed with 70% ethanol, dried, and dissolved in 300 µL TE. Then, RNase was added at a final concentration of 0.1 mg/mL and incubated at 37°C for 30 min, and then an equal volume of a 20% polyethylene glycol 6000 solution containing 2.5 M NaCl was added. After incubation for 30 min on ice, DNA was pelleted by centrifugation, rinsed and then dissolved in TE.

***Meta sequence analysis of bacterial 16S rRNA gene***

The V1-2 region of bacterial 16S rRNA gene (27fmod-338r) was sequenced according to the previous report[18]. The amplified fragments were sequenced on an Illumina Miseq according to the manufacturer's instructions. The paired-end reads were merged using the fastq-join program based on overlapping sequences. Reads with an average quality value of <25 and inexact matches to both universal primers were filtered out. Filter-passed reads were used for further analysis after trimming off both primer sequences. For each sample, quality filter-passed reads were rearranged in descending order according to the quality value and then clustered into OTUs with a 97% pairwise-identity cutoff using the UCLUST program version 5.2.32 (https://www.drive5.com). Taxonomic assignment of each OTU was made by similarity search against the Ribosomal Database Project (RDP) and the National Center for Biotechnology Information (NCBI) genome database using the GLSEARCH program. Indices for α-diversity, namely community richness (Chao1) and diversity (Shannon and Simpson), were calculated, and indices for β-diversity were also estimated using UniFrac analysis with weighted and unweighted principal coordinate analysis (PCoA). All 16S rRNA gene datasets were deposited in GenBank Sequencing Read Achive data base.

***Analysis of fecal metabolites***

All the fecal samples obtained were transported at -20 °C and thereafter stored at −60 °C - −80 °C until analysis. To determine the concentrations of SCFAs (acetate, propionate, butyrate), lactate, and succinate, the contents of frozen fresh fecal samples were determined by using an HPLC Prominence instrument equipped with an electric conductivity detector (CDD-10A$_{VP}$) (Shimazu, Japan)[19] according to the manufacturer's protocol with some modification. In brief, fecal samples (200-400mg) were mixed with 9 fold volume of Milli-Q water for 10 min. After centrifugation at 15,000 rpm, the supernatant were filtrated with 0.45 µm Millex-HA filter (SLHA033SS) (Millipore, USA). The filtrated solutions were prepared for the analysis of the HPLC instrument. The HPLC analysis was carried out on an ion-exclusion column (Shim-pack SCR-102H) (Shimazu, Japan). The measurement conditions were adjusted as follow: mobile phase, 5mM *p*-toluenesulfonic acid; buffer, 5mM *p*-toluenesulfonic acid, 20mM Bis-Tris, and 0.2mM EDTA-4H; setting temperature in the instrument, 40°C; flow rate, 0.8 ml/min.

The other water soluble metabolites of feces were determined by chromatography-tandem mass spectrometry (GC/MS/MS) platforms according to the method described by Nishiumi et al.[1] with some modification. The lyophilized cacel feces were disrupted with zirconia beads using Micro Smash MS-100 (Tomy Seiko, Tokyo, Japan). The resulting freeze-dried 5 mg fecal samples were suspended in 150 µL Milli-Q water containing internal standard (100 µmol/L 2-isopropylmalic acid), and then 150 µL methanol and 60 µL chloroform were added. These samples were homogenized and incubated at 37°C for 30 min with shaking at 1,200 rpm. After centrifugation at 16,000 × g for 5 min at room temperature, 250 µL of the supernatant were transferred to a new tube and added 200 µL of Milli-Q water. After being mixed, the solution was centrifuged at 16,000 × g for 5 min at room temperature, and 250 µL of the supernatant were transferred to a new tube. Samples were evaporated into dryness using a vacuum evaporator system (CentriVap Centrifugal Vacuum Concentrator) (LABCONCO) for 20 min at 40°C and lyophilized using a freeze dryer (TAITEC). Dried extracts were firstly methoxymated with 40 µL of 20 mg/mL methoxyamine hydrochloride (Sigma-Aldrich) dissolved in pyridine. After adding the derivatization agent, samples were shaken at 1,200 rpm for 90 min at 30°C. Samples were then silylated with 20 µL of N-methyl-N-trimethylsilyl-trifluoroacetamide (MSTFA) (GL Science) for 30 min at 37°C with shaking at 1,200 rpm. After derivatization, samples were centrifuged at 16,000 × g for 5 min at room temperature, and the supernatant transferred to glass vial for GC/MS/MS measurement. GC/MS/MS analysis was carried out on a Shimadzu GCMS-TQ8030 triple quadrupole mass spectrometer (Shimadzu) with a capillary column (BPX5) (SGE Analytical Science). The GC oven was programmed as follow: 60 °C held for 2 min, increased to 330 °C (15°C/min), and finally 330°C held for 3.45 min. GC was operated in constant linear velocity mode set to 39 cm/sec. The detector and injector temperatures were 200°C and 250°C, respectively. Injection volume was set at 1 µL with a split ratio of 1:30. Data processing was performed using LabSolutions Insight (Shimadzu).

***Correlation analyses***

The relative values of fecal metabolites dependent upon animal species were analyzed through construction of a correlation heatmap after Spearman correlation coefficient was calculated among fecal metabolites. The fecal metabolites showing significant differences ($p<0.1$) between targeted animal species were selected if with the difference in at least one species. Correlation network analysis were performed as previously described[20]. As the conditions of this study, the metabolites showing high correlation coefficients ($r > |0.4|$) between metabolites were discriminated from each cluster. The data were sorted to 0 ($r<|0.4|$ or 1 ($r >|0.4|$) by using the function "ifelse" of R software (ver. 4.0.5) (https://www.r-project.org/). The plus and minus data as the r value were prepared for visualization of positive and negative correlation network, respectively. The calculated data (.dl file) were rendered as correlation networks by Force Atlas with Noverlap in Gephi 0.9.2 (http://gephi.org).

***Association analyses***

Association analysis, a elementary way of unsupervised learning, is a technically established method, which is also used for market research as market basket analysis, and is applied to understand beyond the logic of numbers, using relative numbers [21-23]. It is easy to apply when there are missing values as a characteristic of association analysis. Therefore, it can be suitable when a condition is a set such that the identity for classification is different for each layer, such as microbial structure and metabolomic data. It is able to obtain and classify predictive associated components by applying it to conditions that are difficult to make horizontal comparisons. In order to predict the components associated with compost and/or thermophile, an association analysis was performed as previously reported [21-23]. In brief, the association analysis is a elementary way to infer the effect, "target", from the cause, "source". In the case, "source" and "target" are represented as the x and y, respectively, calculation factors of association analysis were defined as follows:

support (x ⇒ y) = P(x ∩ y)

"support" is P(xy), joint probability (P) of co-occurance

confidence (x ⇒ y) = P(x ∩ y) / P(x)

"confidence" is P(xy) / P(x), conditional probability (P) of occurence of y after x is occurred.

lift (x ⇒ y) = P(x ∩ y)/P(x) P(y)

"lift" is P(xy) / P(x)P(y), measure of association / independence

Value of > 1 represents positive association (If the value represents independent, value with < 1 represents as negative association).

Here association rules were determined by using criterion values of support, confidence, and lift ( "support = 0.063, confidence = 0.25, maxlen = 2" and "lift > 1.2"). Data jointed with all information such as animal species, phyla, genera, metabolites, with compost and thermophile and without them were used. To avoid the difference dependent upon animal species, all data were calculated based on the median value (M) among the data of the same animal and sorted to 0 ( < M) and 1 ( > M). Therefore, potential associated components should be explored by this analysis. These data set was ruled by the package "arules" of R software. The systemic network was rendered by Force Atlas with Noverlap in Gephi 0.9.2.

***Energy landscape analysis***

Energy landscape analysis (ELA) was performed as a machine learning method as previously described[24]. ELA is a data-driven method for constructing landscapes that explain the stability of communities compositions across environmental gradients. Here, ELA is based on an extended pairwise maximum entropy model that explains the probability for the occurrence of the ecological state of sample $k$, $\sigma^{(k)}$ given the environmental condition $\epsilon^{(k)}$; as the ecological state, we combined the presence/absence status of selected taxa and levels of physiological factors,

$$\sigma^{(k)} = \left(\sigma_1^{(k)}, \sigma_2^{(k)}, \dots \sigma_N^{(k)}\right) = \left(\sigma_1^{(k)}, \dots, \sigma_{n_m}^{(k)}, \sigma_{N-n_c}^{(k)} \dots \sigma_N^{(k)}\right)$$

where $n_m$ is the number of bacterial taxa and $n_c$ is the number of metabolic chemicals; as the environmental condition, two environmental factors representing with (1) or without (0) compost-derived thermophile, ($\epsilon_i^{(k)}$) and host (fish, chicken, pig, cattle) converted to the 0-1 range ($\epsilon_i^{(k)}$; compost-derived thermophile are converted as  1, respectively; ) were combined as $\epsilon^{(k)} = (\epsilon_i^{compost}$, $\epsilon_i^{fish}, \epsilon_i^{chicken}, \epsilon_i^{pig}, \epsilon_i^{cattle})$. The model can be written as:

(I) $$P(\sigma^{(k)}|\epsilon\ ) = \frac{e^{-E(\sigma^{(k)}|\epsilon\ )}}{\Sigma e^{-E(\sigma^{(k)}|\epsilon\ )}},$$

(II) $$E\left(\sigma^{(k)}\middle|\epsilon^{(k)}\right) = -(\Sigma_i \Sigma_j J_{ij} \sigma_i^{(k)} \sigma_j^{(k)} + \Sigma_i g_i^a \epsilon_{a,i}^{(k)} \sigma_i^{(k)} + \Sigma_i g_i^s \epsilon_{s,i}^{(k)} \sigma_i^{(k)} + \Sigma_i h_i \sigma_i^{(k)}).$$

Here, $P(\sigma^{(k)}|\epsilon)$ is the probability for the occurrence of an ecological state $\sigma^{(k)}$. Eq. (I) shows that the probability is high when energy $E\left(\sigma^{(k)}\middle|\epsilon\right)$ is low and *vice versa*. In eq. (II), $E\left(\sigma^{(k)}\middle|\epsilon\right)$ is defined as the sum of the effect of interaction among components, antibiotic treatment, growth stages, and the net effect of unobserved environmental factors. Parameters in eq. (II), namely, $J_{ij}$, $g_i^a$, $g_i^s$ and $h_i$, indicate the effect of the relationship among components ($J_{ij} > 0$ favors and $J_{ij} < 0$ disfavors the cooccurrence of components $i$ and $j$), the effect of the antibiotics on component $i$ (the antibiotic treatment positively ($g_i^a > 0$) or negatively ($g_i^a < 0$) affects the occurrence of component $i$), the effect of the growth stages on component $i$ (component $i$ favors the later ($g_i^s > 0$) or early ($g_i^s < 0$) growth stage) and how likely component $i$ occurs when the other factors are equal, respectively. To obtain p-values for each parameter, a permutation test [25] was used as 2,000 stimulations. All the components in $\sigma^{(k)}$ were converted to the 0-1 range as follows. We first interpret the relative abundance of each bacterial genus in the samples as the area close to the facility (No.31 in Fig. S6) (1) or 100 m away (No.31_100 m in Fig. S6) (0) states by setting a threshold value of 0.001. Then, we selected 51 families ( >1% as the maximum value of the bacterial population in all seasons). Consequently, we obtained the set of explanatory variables $\sigma^{(k)}$ with 51 components, which accompanies the environmental condition $\epsilon^{(k)}$ represents the status of the distance from the aquaculture facility and the seasonal stage.

### ***Structural Equation Modeling***

Structural equation modeling (SEM) for confirmatory factor analysis (CFA) was conducted using the package "lavaan"[26,27] of R software. The analysis codes were referred to the website (https://lavaan.ugent.be). Since CFA needs a hypothesis, the groups selected by association analysis were utilized as factors for a latent construct of metabolites and microbiota. The models as hypotheses were statistically estimated using maximum likelihood (ML) parameter estimation with bootstrapping (n = 1000) by the functions 'lavaan' and 'sem'. Model fit was assessed by the chi-squared p-value (p >0.05, nonsignificant), comparative fit index (cfi) (>0.9), Tucker–Lewis index (tli) (>0.9), goodness-of-fit index (gfi) (>0.95), adjusted goodness-of-fit index (agfi) (>0.95), and root mean square error of approximation (rsmea) (<0.05) as indices of good model fit.[28] The path diagrams of the good model were visualized with layout="spring" using the package "semPlot" of R software [29].

***In vitro virus infection test***

Virus infectivity control tests were conducted with reference to previous reports[30,31]. In brief, as preliminary test, the effect of the test material on cultured cells (cytotoxicity) was investigated. As the materials, 1/100 diluted compost extract, the vegetative cell of Caldibacillus hisashii itself (at least $10^6$ cfu/ml) and the derived culture solution. The test materials were inoculated into cultured cells after 10-fold step dilution in phosphate buffer solution, and the highest concentration at which the cells showed normal condition after incubation was confirmed to determine the virus concentration to be used for the test. As a result, no cytotoxicity was confirmed in the 10-fold dilution. Therefore, the detection limit in this study was set at 101.5 TCID50 / mL. Ten mL of each of the test materials and phosphate buffer were aliquoted, and 1 mL of the virus solution was added to the concentration determined in the preliminary test, then incubated at 25°C for 3 hours. The cell lines suitable for each of virus to test were maintained in Dulbecco's modified Eagle's medium supplemented with 10% FBS and glutamine–penicillin–streptomycin solution in 5% $CO_2$ in humidified air at 37 °C. The targeted virus was inoculated into a monolayer of the suitable cell lines in the 96 well plates. The plates were incubated at 37°C in 5% $CO_2$ for 5 days and cytopathic effect (CPE) in each well was observed. Each cell line is shown as follows: Vero cell line (African green monkey kidney epithelial-derived strain cell line) for *Alphacoronaviurs* (PED virus, Porcine epidemic diarrhea virusP-5V strain) and *Betacoronavirus* (SARS-CoV-2, δstrain); and MDCK cell line（Canine kidney-derived strain cell line）for *Influenzavirus*(swine influenza virus H1N1 IOWA strain). SARS-CoV-2 , human-derived isolates were used. : After isolation and culture using Vero cells from saliva, real-time PCR was used to confirm amplification of the SARS-CoV-2 gene (Ministry of Health, Labor and Welfare notification method) and to confirm changes in N501Y (-), L452R (+) virus strains were used. These approvals and experiments were performed by Shoku-Kan-Ken, Inc. (Laboratory of Food Environment and Hygience), Gumma, Japan.

***Germ free mouse experiment***

Oral administration test with germ free mice were conducted by the methods on the basis of previous reports[8,32]. In brief, BALB/c germfree male mice were used in accordance with the guidelines for the care and use of laboratory animals at Yokohama City University. The germ-free mice (n = 5)(8-week-old) received either autoclaved water with or without the vegetative cells (including spore) of *C. hisashii*. The mice received water *ad libitum* for the experiment.

***Protein structure prediction***

After converting the target gene sequence into an amino acid sequence in the genome sequence data of *Caldibacillus hisashii* N11 strain, prediction of peptide structure was performed by Alphafold 2 according to the developed procedure (https://alphafold.ebi.ac.uk/faq)[33,34].

***Statistical analyses***

Data for intestinal bacterial flora and fecal metabolites concentrations were analyzed by F test and Kolmogorov-Smirnov test to select parametric and non-parametric analyses. A paired t-test, Wilcoxon test, ANOVA followed by Tukey's post hoc test, and Kruskal-Wallis one-way analysis of variance followed by Steel-Dwass test as appropriate methods dependent upon data sets were performed. Kaplan–Meier curves was statistically compared by log-rank test. Significance was declared when $P < 0.05$, and tendency was assumed at $0.05 \leq P < 0.20$. These calculation data were prepared by using the R software and Prism software (verstion9.1.2), which was also visualized with Entering replicate data graph. Data in pathway analysis was calculated and visualized by using MetaboAnalyst 5.0[35-37]. Data are presented as the means ± SE.

**Data availability.** Raw files of the bacterial V1–V2 16S rRNA data are deposited in the DNA Data Bank of Japan (DDBJ), under NCBI Bio-Project accession numbers PRJDB9535 (PSUB012136) . The 16S rDNA library in Extended fig.1 are also deposited in the DDBJ, under NCBI Bio-Project accession numbers LC646904-LC646942.